\documentclass[aps,pre,twocolumn,superscriptaddress]{revtex4}

\usepackage{graphicx, color}
\usepackage{amssymb,amsfonts,amsmath}
\usepackage{tabulary}
\usepackage{booktabs}

\def\s{\sigma}

    \def \a{\alpha}
    \def \b{\beta} 
\def \s{\sigma}      
\def \e{\epsilon}    
   \def \d{\delta} 
    \def \l{\lambda}

\newcommand{\matr}[1]{\mathbf{#1}} 
\newcommand{\bvec}[1]{\boldsymbol{#1}}

\def \del{\partial}    % for writing partial derivatives 

\def \HF{\dfrac{1}{2}}  % small and big half's. 

\def \>{\rangle} 
\def \<{\langle} 
 
\def\be{\begin{equation}} 
\def\ee{\end{equation}} 
\def\longrightharpoonup{\relbar\joinrel\rightharpoonup}
\def\longleftharpoondown{\leftharpoondown\joinrel\relbar}

\def\longrightleftharpoons{
  \mathop{
    \vcenter{
      \hbox{
      \ooalign{
        \raise1pt\hbox{$\longrightharpoonup\joinrel$}\crcr
	  \lower1pt\hbox{$\longleftharpoondown\joinrel$}
	  }
      }
    }
  }
}

\newcommand \bea {\begin{eqnarray}} 
\newcommand \eea {\end{eqnarray}}

\begin{document}

\title{Habitat Fluctuations Drive Species Covariation in the Human Microbiota } 

\author{Charles K. Fisher}
\email{fisher@lpt.ens.fr}
\affiliation{Laboratoire de physique th\'eorique, CNRS, UPMC and \'Ecole normale sup\'erieure, 24 rue Lhomond, 75005 Paris, France}

\author{Thierry Mora}
\affiliation{Laboratoire de physique statistique, CNRS, UPMC and \'Ecole normale sup\'erieure, 24 rue Lhomond, 75005 Paris, France}

\author{Aleksandra M. Walczak}
\affiliation{Laboratoire de physique th\'eorique, CNRS, UPMC and \'Ecole normale sup\'erieure, 24 rue Lhomond, 75005 Paris, France}
\date{\today}

\begin{abstract}
Two species with similar resource requirements respond in a characteristic way to variations in their habitat -- their abundances rise and fall in concert. We use this idea to learn how bacterial populations in the microbiota respond to habitat conditions that vary from person-to-person across the human population. Our mathematical framework shows that habitat fluctuations are sufficient for explaining intra-bodysite correlations in relative species abundances from the Human Microbiome Project. We explicitly show that the relative abundances of phylogenetically related species are positively correlated and can be predicted from taxonomic relationships. We identify a small set of functional pathways related to metabolism and maintenance of the cell wall that form the basis of a common resource sharing niche space of the human microbiota. 
 \end{abstract}
\maketitle

Species in an ecosystem interact with each other and with their environment. Both types of interactions leave an imprint on the composition and diversity of a community. Two species competing for exactly the same resources engage in a struggle for existence \cite{macarthur_limiting_1967}. In the end, one species will win the competition by driving the other to extinction. As a result, one might expect that closely related species rarely occupy the same habitat where they would risk being drawn into competition. On the other hand, species that survive in the same habitat must share many common features \cite{chesson2000mechanisms}. Thus, the rise and fall of a common resource may cause the abundances of similar species to rise and fall in concert. These opposing ecological forces simultaneously push and pull on species abundances to shape the composition of a community. 

Ecological processes operate on the thousands of microbial species that inhabit the human body \cite{turnbaugh2007human, human2012framework, human2012structure, dewhirst2010human} just as they operate on the Amazon rainforest. Technological advances have recently made it possible to study the human microbiota using 16S ribosomal RNA tag-sequencing and whole genome `shotgun' metagenomics \cite{wooley2010primer}. Variability in the composition of the microbiota can be studied in two ways. Longitudinal studies follow the relative abundances of the species in a single bodysite of a particular person over time \cite{caporaso2011moving}. Cross-sectional studies examine the relative abundances of the species in a single bodysite across a sample of many different people \cite{yatsunenko2012human}. These studies have demonstrated that the composition of the human microbiota exhibits three scales of variation \cite{costello2009bacterial,claesson2011composition}: there are small-scale fluctuations in relative species abundances through time, there are medium-scale variations in species composition from person-to-person, and there are large-scale differences in species composition between different bodysites. 

\begin{figure*}
\centering
\includegraphics[width=\textwidth]{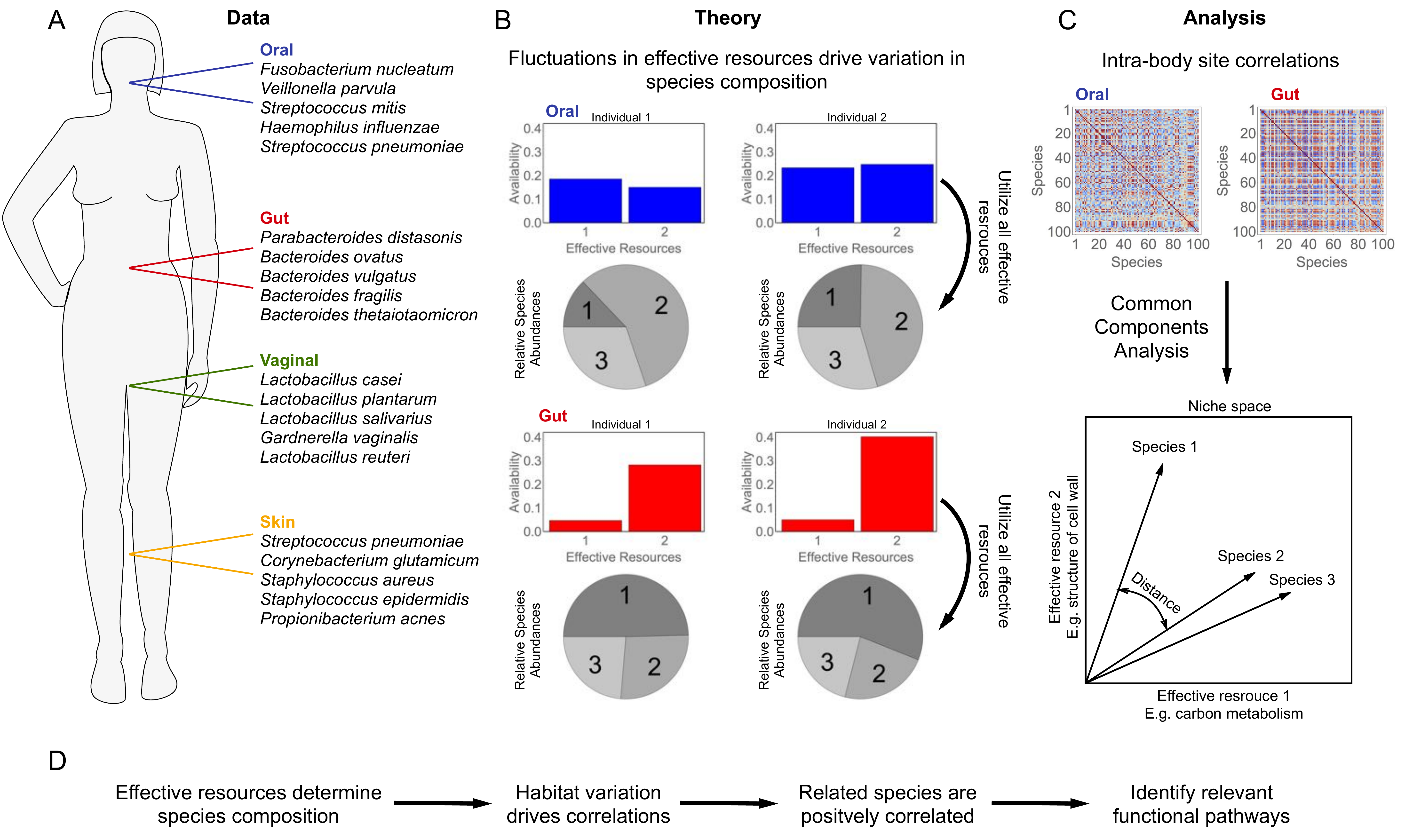}
\caption{ \textbf{ Schematic outline of the paper. }  A) Data on the microbial species composition of the gut, oral cavity, vagina, and skin from the Human Microbiome Project were obtained from MG-Rast. One hundred highly abundant species were selected for further study (Supporting Information). The top five most abundant species in each bodysite are shown. B) We propose a theory for the composition of the microbiota based on a Principle of Maximum Diversity, which says that the equilibrium relative species abundances in a community maximize the diversity of the community while ensuring that all niches are fully utilized. As a consequence, fluctuations from person-to-person in the species composition within a bodysite result from fluctuations in the availabilities of some `effective resources' that define the space of niches. C) The theory predicts that the intra-bodysite variations in the relative abundances of species occupying similar regions of niche space will be correlated. Therefore, we can infer the structure of the niche space of the human microbiota from intra-bodysite correlations between species using a technique called Common Components Analysis (Supporting Information). D. Overall logical flow of the paper. Characterizing the effective resources allows us to identify the functional pathways which must be conserved to utilize a common habitat.}
\label{fig:fig1}
\end{figure*}

Previous studies have interpreted correlations in relative species abundances from cross-sectional studies as effective interactions between species \cite{friedman2012inferring, kurtz2014sparse, segata2013computational}. However, simulations with classical ecological models \cite{fisher2014identifying} demonstrate that the correlations between species abundances observed in cross-sectional studies do not reflect actual competitive or mutualistic relationships between species. By contrast, we demonstrate in this work that {\it habitat variability} is sufficient to explain the medium-scale variations in species composition observed in cross-sectional studies of the human microbiota. As a result, the relative abundances of phylogenetically related species are positively correlated -- they rise and fall in concert as habitat conditions vary from person-to-person. Therefore, cross-sectional studies allow us to extract a wealth of information about the influence of species traits and habitat properties on community composition using advanced statistical techniques. 

We analyzed data from the Human Microbiome Project (HMP) on person-to-person variability in relative species abundances in four bodysites (gut, oral cavity, vagina, and skin; Figure \ref{fig:fig1}a) \cite{turnbaugh2007human,human2012structure,human2012framework,gevers2012human}. The data were analyzed using a mathematical model formulated with the assumption that all intra-bodysite variability in species composition is driven by varying habitat conditions (Figure \ref{fig:fig1}b; Supporting Information). Specifically, we say that each sample from a bodysite reflects a habitat containing a variety of effective resources that reflect all of the abiotic and biotic factors that influence species composition in the community. We propose that the relative species abundances can be determined from the conditions of the habitat using a Principle of Maximum Diversity, which says that the equilibrium relative species abundances in a community maximize diversity while ensuring that all effective resources are fully utilized (see Supporting Information). Mathematically, this is expressed in an equation of the form:
\begin{widetext}
\be
\log ( \text{abundance} ) = \text{constant} + \sum_{ \text{resources} } (\text{availability of resource}) \times ( \text{ability to use resource} ) \nonumber
\ee
\end{widetext}
The availabilities of the effective resources vary from person-to-person, causing the relative abundances of the species to vary as well. As a result, the relative abundances of species that use similar effective resources will be correlated, and it is possible solve an inverse problem to learn which species use which effective resources from these correlations (Figure \ref{fig:fig1}c-d). 

\begin{figure*}
\centering
\includegraphics[width=\textwidth]{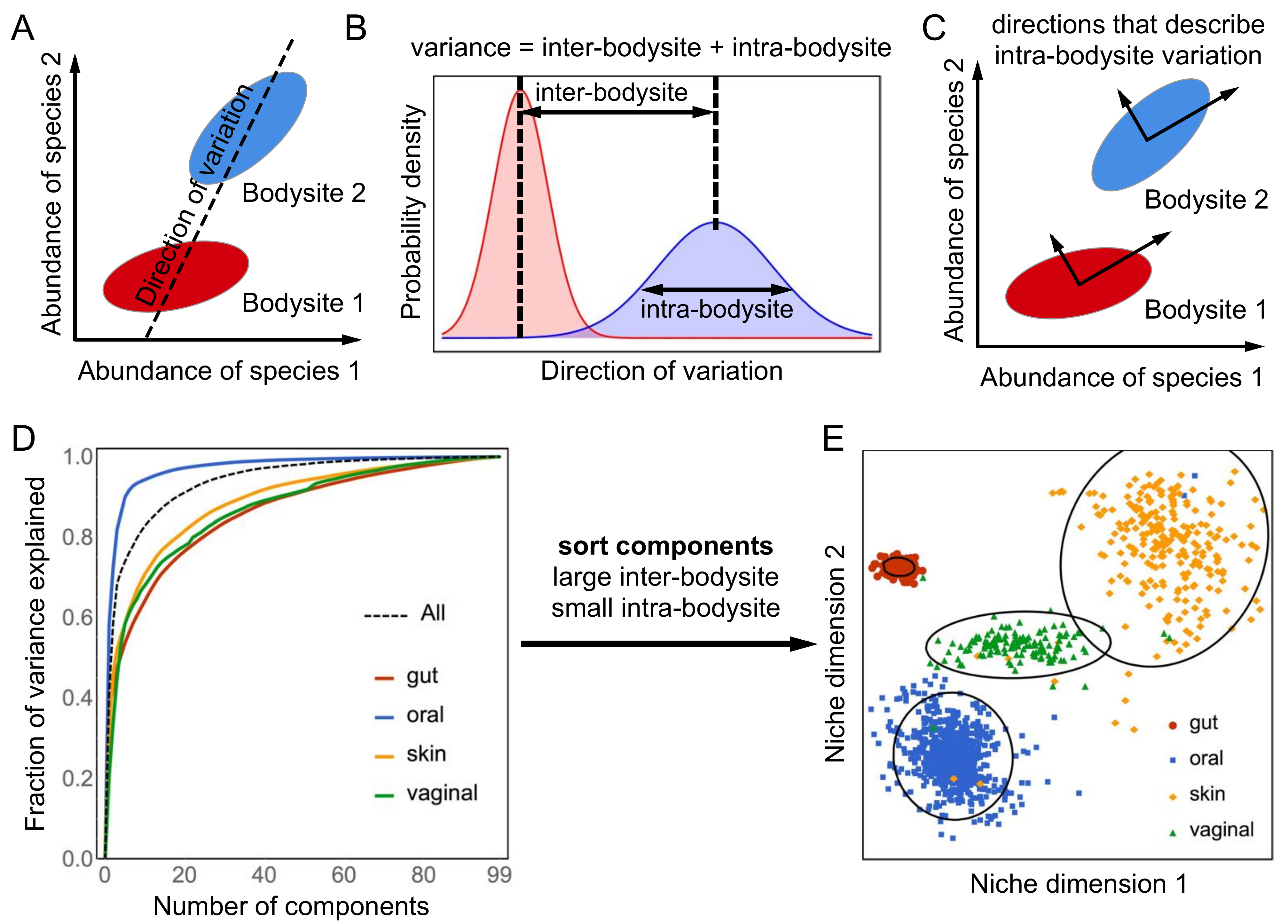}
\caption{ \textbf{Describing intra-bodysite variability with common components analysis.}  A) Covariation of the relative abundances along any direction captures both inter-bodysite differences and intra-bodysite variability. B) The law of total variance states that the variance along a direction is $\text{variance} = \left( \text{inter-bodysite differences} \right)^2 + \text{intra-bodysite variance}$. C) Common Components Analysis (CCA) finds a common set of directions that simultaneously capture intra-bodysite variability in each of the four bodysites. D)  Most of the intra-bodysite variability in species composition can be explained with a small number of common components. E) Projecting onto two common components with large inter-bodysite differences and small intra-bodysite variation clearly separates the four body sites.}
\label{fig:fig2}
\end{figure*}

The inverse problem can be solved because, by construction, the model imposes that the inferred effective resources correspond to directions with high intra-bodysite variability (Figure \ref{fig:fig2}a-c). We exploit this feature to developed a technique called Common Components Analysis (CCA) that infers the characteristics of the species and habitats from observed correlations (see Supporting Information). CCA aims to find a single set of directions that simultaneously explain variation within each of the bodysites. For comparison, another common data analysis technique called Principal Components Analysis (PCA) aims to find a set of directions that explain total variability \cite{dunteman1989principal}, which is a sum of inter-bodysite differences and intra-bodysite variability. Our maximum diversity model suggests that it is not possible to learn which species use which resources from the inter-bodysite differences. Therefore, we hypothesize that CCA will provide a better description of the HMP data than PCA.

We applied CCA to study the relative abundances of $N = 100$ highly abundant species from the HMP (see Supporting Information). To validate the mathematical model underlying CCA, we verified that its assumptions, which are rooted in our hypothesis that compositional variation is driven by habitat fluctuations, are not violated (Supporting Information). Fitting the data while fulfilling all of the modeling assumptions is not guaranteed. Indeed, the method fails when applied to randomized correlation matrices (Figure S1-S5), which demonstrates that the observed performance is not due to chance. Figure \ref{fig:fig2}d shows that CCA identifies a few components (i.e., effective resources) that explain most of the intra-bodysite variation in the human microbiota. Even though the CCA components only capture the directions that explain intra-bodysite variability, they can be ranked by the ratio of how much they vary between bodysites to how much they vary within bodysites. Sorting the CCA components in this way identifies directions that clearly separate all four bodysites into coherent clusters (Figure \ref{fig:fig2}e). By contrast, the principal components are typically ranked based on their contribution to total variability, which is a mixture of inter- and intra-bodysite variation. Thus, highly ranked principal components may correspond to directions with large intra-bodysite variations, causing them to miss directions with large inter-bodysite differences. As a result, the two largest principal components are unable to separate all four bodysites (Figures S6 and S7). 

\begin{figure*}
\centering
\includegraphics[width=\textwidth]{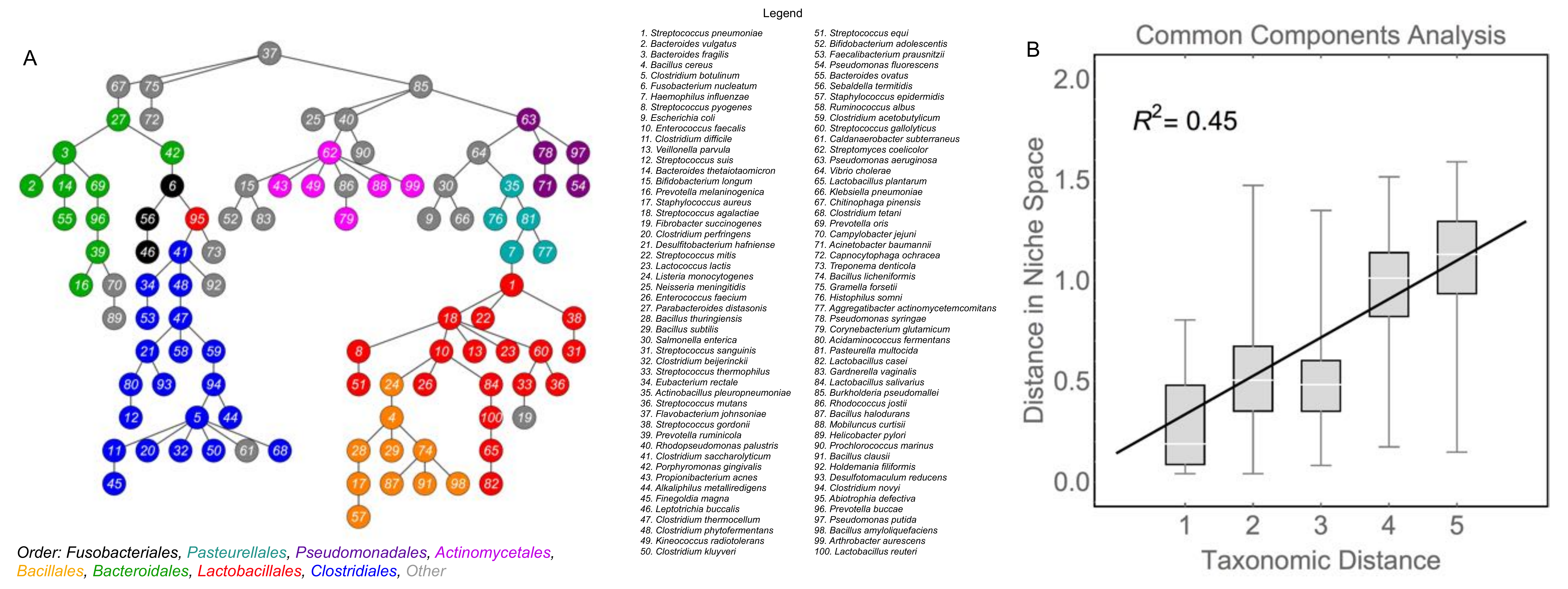}
\caption{ \textbf{Visualizing the niche space of the human microbiota.} Each species can be described as a vector within a niche space that describes the ability of a species to utilize the effective resources, weighted by the average variability of that resource within each of the bodysites. Distances between two species in this niche space were quantified with a metric that measures the angle the two vectors (Supporting Information). A) The minimum spanning tree of the niche space connects all of the species so that the average distance between connected species is as small as possible. Thus, two species are connected if they have similar effective resource utilizations. The species are colored according to their taxonomic classification at level `order'. Only the top eight orders with the most representative species are colored; species in underrepresented orders are shown in gray. C) The distance between species in the niche space obtained from CCA is strongly related to species similarity. Here, the taxonomic distance between two species is five minus the number of overlapping taxonomic levels. See Supporting Information for discussion of statistical significance.  }
\label{fig:fig3}
\end{figure*}

CCA describes each species as a vector, where each element of the vector describes the ability of that species to use one of the effective resources. Thus, the angle between two of these vectors describes how similar the two species are in terms of their abilities to use the resources. Species that are positively correlated are close together in this `niche space', whereas species that are uncorrelated (or anti-correlated) are far apart. Figure \ref{fig:fig3}a shows all of the species connected into a tree, so that each species is only connected to other species with similar effective resource utilizations. Coloring the tree by taxonomic classification at the level of `order' reveals that the species cluster into taxonomically coherent groups \cite{federhen2012ncbi}. In fact, the more similar two species are in terms of taxonomy, the closer they are in this niche space (Figure \ref{fig:fig3}b; Figures S5 and S7). To put it another way, the relative abundances of related species are highly correlated because they have similar resource requirements. This is true even though species that use similar resources are competing with each other. The intuition derived from dynamical models that the abundances of competing species should be anti-correlated simply does not apply when the habitat conditions are highly variable. 

\begin{figure*}
\centering
\includegraphics[width=\textwidth]{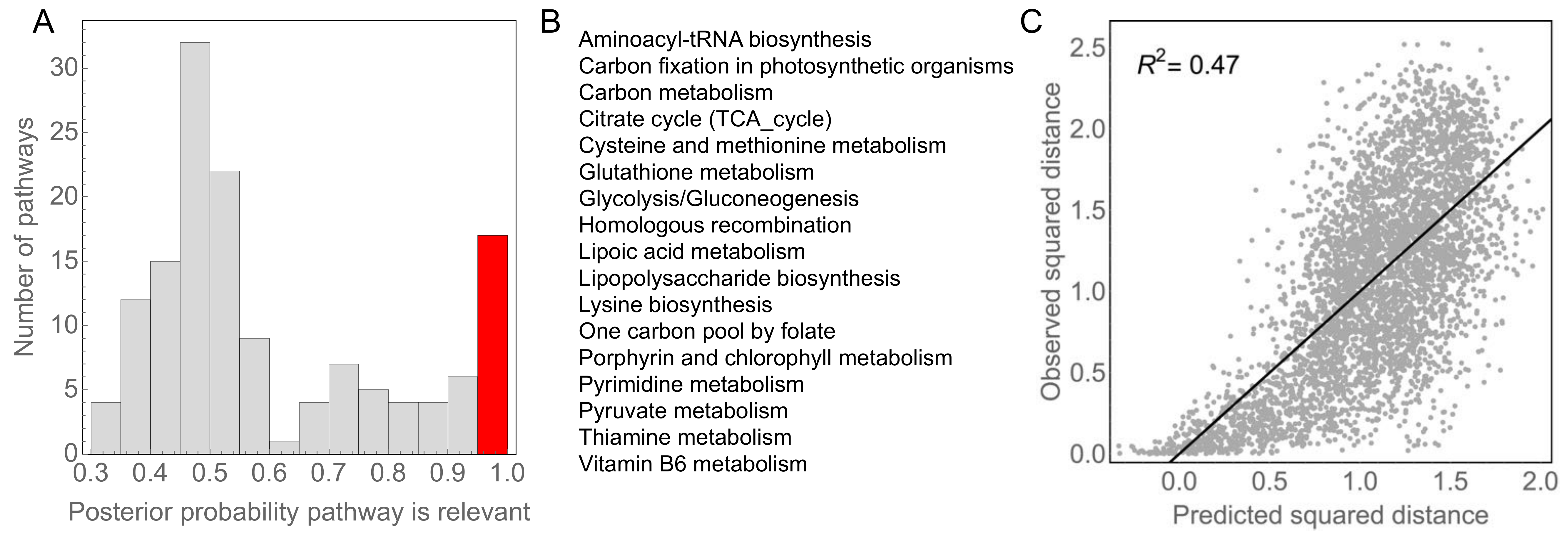}
\caption{ \textbf{Functional pathways related to inter-species distances in niche space.} We performed a linear regression of the squared distances computed from CCA against the squared distances computed from KEGG functional pathways (Supporting Information). Each pathway was assigned a probability that it contributes to the distances between species in niche space (i.e. that the regression coefficient associated with the pathway is nonzero) using a Bayesian model selection algorithm (Supporting Information).  A) A histogram of the probabilities associated with each of the KEGG pathways. We defined pathways as significantly associated if they had a posterior probability greater than $0.95$, and the bar representing the significant pathways is shown in red. B) A list of the relevant function pathways in alphabetical order. C) The regression using just these relevant pathways has a correlation coefficient of $R^2 = 0.47$. }
\label{fig:fig4}
\end{figure*}

So far, we have described the components derived from CCA as abstract resources that represent unknown abiotic and biotic factors in the habitat. To check that these effective resources correspond to biologically meaningful functions, we regressed the distances between species in the CCA derived niche space against inter-species distances computed from KEGG functional pathways (see Supporting Information) \cite{kanehisa2000kegg, mazumdar2013metabolic}. We used a statistical technique called the Bayesian Ising Approximation to assign a posterior probability to each KEGG pathway \cite{fisher2015bayesian,fisher2014bayesian} (Supporting Information). The posterior probability is a measure of degree of belief; it quantifies how relevant each KEGG pathway is for computing the similarity between species derived from CCA. A histogram of the posterior probabilities is shown in Figure \ref{fig:fig4}a (also Figures S8 and S9). We designated pathways as relevant if they had a posterior probability greater than 0.95. The 17 pathways reaching this threshold for relevance are listed in alphabetical order in Figure \ref{fig:fig4}b. Taken together, these relevant pathways explain roughly half of the variation in the distances between species in the CCA derived niche space (Figure \ref{fig:fig4}c).  The relevant pathways are primarily associated with carbon metabolism or maintenance of the cell wall, pointing to the importance of both metabolic processes and host-microbiome interactions for structuring the human microbiota. 

Previous studies have revealed that bacteria exhibit tremendous genomic and functional diversity due, in part, to high rates of horizontal gene transfer (HGT) \cite{smillie2011ecology}. As a result, the ability of sequence-based or taxonomic classification of bacteria to capture ecological relationships has been called into question \cite{tikhonov2015interpreting, koeppel2013surprisingly, philippot2010ecological, doolittle1999phylogenetic}. Nevertheless, we found that genetically related species respond to fluctuating habitat conditions in the same way, implying that they occupy similar ecological niches. Thus, current taxonomic groupings of bacteria are largely sufficient for explaining cross-sectional correlations in relative species abundances over the healthy human population. This result is not at odds with high rates of HGT; it simply implies ecologically derived constraints on evolution. The methods developed here (e.g., CCA) can be applied to any cross-sectional study with labeled metadata, including studies with populations corresponding to healthy and unhealthy individuals. Given that CCA identifies features that separate the bodysites with high fidelity, we believe that it is a useful technique for identifying microbiota based biomarkers that discriminate between host phenotypes. Extending our results to include data from unhealthy subjects will be an important avenue for future work.

\section{Appendix}
\setcounter{equation}{0}
\setcounter{figure}{0}
\makeatletter 
\renewcommand{\theequation}{S\@arabic\c@equation} 
\makeatletter 
\renewcommand{\thefigure}{S\@arabic\c@figure} 
\makeatletter 
\renewcommand{\theequation}{S\@arabic\c@equation} 

\appendix

%\tableofcontents
\begin{table*}[htbp]\caption{Summary of Frequently Used Symbols}
\centering % to have the caption near the table
\begin{tabular}{r c p{10cm} }
\toprule
$s$ & $:=$ & Index for the bodysites (e.g., the gut)\\
$N$ & $:=$ & Number of species\\
$i$ & $:=$ & Index for species\\
$M$ & $:=$ & Number of effective resources. Set to $M = N-1$\\
$\mu$ & $:=$ & Index for effective resources\\
$V_{i \mu}$ & $:=$ & Utilization of effective resource $\mu$ by species $i$\\
$\lambda_{\mu}$ & := & Lagrange multiplier that encodes availability of resource $\mu$\\ 
$\bvec{x}$ & $:=$ & Vector of relative abundances\\
$\matr{G}$ &$:=$ & $(N-1) \times N$ matrix that implements the additive log-ratio transform\\
$\tilde{\matr{V}}$ &$:=$& $\matr{G} \matr{V}$: Transformed resource utilization matrix\\ 
$\bvec{y}$ &$:=$& $\matr{G} \log \bvec{x}$: Vector of additive log-ratio transformed relative abundances\\
$\matr{\Psi}_s$ &$:=$& Empirical covariance matrix of additive log-ratio transformed relative abundances in bodysite $s$\\
$\matr{\Psi}$ &$:=$& Empirical covariance matrix of additive log-ratio transformed relative abundances across all bodysites\\
$\matr{\Sigma}_s$ &$:=$& Covariance matrix of $\bvec{\lambda}$ in bodysite $s$\\
$\bar{\bvec{\lambda}}_s$ &$:=$& Average of $\bvec{\lambda}$ in bodysite $s$\\
\bottomrule
\end{tabular}
\label{tab:TableOfNotationForMyResearch}
\end{table*}

\renewcommand{\thefigure}{S\arabic{figure}}
\renewcommand{\theequation}{S\arabic{equation}}

\section{Primer on Compositional Data Analysis }

\subsection{ Metagenomic data on species abundances are compositional }

Metagenomic studies of species abundances provide data in the form of counts. That is, the number of sampled sequences assigned to species $i = 1, \ldots, N$ is given by the integer count $n_i$.  We assume the counts are proportional to the true species abundances, but the constant of proportionality (which may vary between experiments) is unknown. As a result, we follow the standard procedure by analyzing the relative species abundances \cite{friedman2012inferring}:
\be
x_i = \frac{ n_i }{ \sum_j n_j}
\ee
Relative species abundances are \emph{compositional}, which means that $x_i \geq 0$ and $\sum_i x_i = 1$. The normalization removes a degree of freedom (i.e. we do not know the total population size), and imposes significant restrictions on the types of modeling and analyses that can be performed using relative abundances. Therefore, this section will review the principles of compositional data analysis that will guide our analysis of the microbiota \cite{aitchison2005compositional}. 

\subsection{Working with compositions }

Aitchison introduced a framework for the statistical analysis of compositions in 1982 \cite{aitchison1982statistical}. A compositional vector $\bvec{x}$ of length $N$ lies in the simplex $\mathcal{S}^N = \{ \bvec{x} \in \mathcal{R}_+^N | \sum_i x_i = 1\}$. For example, $x_i$ could be the relative abundance of a species in a community, as described above. According to Aitchison, the fundamental principle of compositional data analysis is \cite{aitchison2008single}:
\begin{quote}
Any meaningful function of a composition can be expressed in terms of ratios of the components of the composition. Perhaps equally important is that any function of a composition not expressible in terms of ratios of the components is meaningless.
\end{quote}
That is, because the total population size is unknown, all theory and methods of analyses may only use ratios of the relative abundances, which satisfy $x_i/x_j = n_i /n_j$ and, therefore, are insensitive to variations in the total population size. Analyzing compositional data in a manner that does not respect this principle may lead to incorrect conclusions, such as attributing spurious correlations between species to improper causes \cite{aitchison1982statistical,friedman2012inferring,lovell2014proportionality,lovell2013have}. 

\subsection{Transforms for compositional data }

In practice, analyzing compositions amounts to computing a transformation of the data of the form $\bvec{y} = \matr{G} \log \bvec{x}$ where $\matr{G}$ is a matrix that maps the vector of 1's to the vector of 0's (i.e. $\matr{G} \bvec{1} = \bvec{0}$) \cite{aitchison2005compositional,lovell2014proportionality}. The mathematics of compositions ensures that performing analyses on the transformed data, instead of directly on the relative abundances, respects the fundamental principle of compositional data analysis. One common choice for $\matr{G}$ is the centering matrix $\matr{C} = \matr{I} - N^{-1} \bvec{1} \bvec{1}^T$, where $\matr{I}$ is the $N \times N$ identity matrix. This implements the `centered log-ratio'  (CLR) transform. In this work, however, we have chosen to use the `additive log-ratio' (ALR) transform, which is obtained using the $(N-1) \times N$ matrix $\matr{G}$ given by:
\[ \matr{G} = \left( \begin{array}{cccccc}
-1 & 1 & 0 & 0 & \cdots & 0 \\
-1 & 0 & 1 & 0 & \cdots & 0 \\
\vdots & \vdots & \ & \ddots & \ & \vdots \\
-1 & 0 & \cdots  & 0 & 1 & 0 \\
-1&  0 & \cdots  & 0 & 0 & 1 \end{array} \right)  \]
The additive log-ratio transform maps a compositional vector of length $N$ to a vector of real numbers of length $(N-1)$ with elements $y_i = \log( x_{i+1} / x_1 )$ for $i = 1, \ldots, N-1$. The decrease in the dimension of the vector reflects the loss of a degree of freedom due to ignorance of the total population size. 

\section{Theoretical Model of Species Composition  }

\subsection{A principle of maximum diversity }

Relative species abundances provide no information about the total population size in a community. Unfortunately, the total population size plays an important role in most theoretical models of population dynamics (so-called `density dependent' effects). Therefore, our goal in this section is to develop a model of relative species abundances using a theory that does not include any explicit density dependent effects. In this section, we present the theory as a generative model in which known environmental conditions and species properties are used to deduce the relative species abundances. In practice, we will use the model in the opposite direction; that is, observed relative species abundances across many different environments will be used to infer properties of the species. We will describe the `inverse model' in the next section.

Rather than starting from an existing model based on absolute abundances, we postulate a fundamental principle acting directly on the relative abundances of the species in a community. The intuition for our model comes from a classic idea in ecology (e.g. MacArthur \cite{mac1969species}) that the equilibrium species composition of a community ensures that every niche is being fully utilized. This idea can be treated explicitly in some consumer-resource models but we will consider a more general model in which the niches are abstract properties of the community \cite{mac1969species,macarthur1970species,chesson_macarthurs_1990}.

Our model for relative species abundances is based on a postulate we refer to as the `principle of maximum diversity':
\begin{quote}
\textbf{Principle of Maximum Diversity:} The equilibrium relative species abundances in a community maximize the diversity of the community while ensuring that all effective resources are fully utilized.
\end{quote}
To formalize the model, we suppose that a particular community has $M$ effective resources that define the dimensions of the niche space. Note that we use the term `effective resources' in an abstract way that captures all of the abiotic and biotic factors that affect the species composition within a community. Each effective resource $\mu$ has an availability $V_{\mu}$, which varies between different environments. It is the variation in the availabilities of the effective resources that drives variation in species composition between communities. In a community composed only of species $i$, an amount $V_{i \mu}$ of resource $\mu$ will be utilized. In other words, $V_{i \mu}$ describes the ability of species $i$ to utilize resource $\mu$. Note that $V_{i \mu}$ can be positive, in which case species $i$ depletes resource $\mu$, or it could be negative, in which case species $i$ adds more of resource $\mu$ to the environment. For example, a bacterium may secrete a metabolite that is utilized by other species. Finally, we quantify the diversity of a community using the Shannon entropy $H[\bvec{x}] = -\sum_i x_i \log x_i $ \cite{shannon2001mathematical,macarthur1955fluctuations}. Following the principle of maximum diversity, the equilibrium relative species abundances can be obtained by maximizing $H[\bvec{x}]$ subject to subject to constraints $V_{\mu} = \sum_i V_{i \mu} x_i$ and $\sum_i x_i = 1$. 

We can solve for the equilibrium relative abundances by maximizing the Lagrangian:
\begin{align}
L(\bvec{x}, \bvec{\lambda}, \gamma) &= -\sum_i x_i \log x_i +  \sum_{\mu} \lambda_{\mu} (V_{\mu} - \sum_i V_{i \mu} x_i) \nonumber \\
&+ \gamma ( 1 - \sum_i x_i)
\end{align}
where $\gamma$ and the $\lambda_{\mu}$'s are Lagrange multipliers. The solution is given by
\be
x_i^* = \frac{ e^{\sum_{\mu} \lambda_{\mu} V_{i \mu} } }{ \sum_j e^{\sum_{\mu} \lambda_{\mu} V_{j \mu} } }
\label{eq:equil}
\ee
where the Lagrange multipliers are chosen so that
\be
\frac{ \del }{ \del \lambda_{\mu} } \log \sum_j e^{\sum_{\mu} \lambda_{\mu} V_{j \mu}} = \sum_i V_{i \mu} x_i^* =  V_{\mu}
\ee
There is one degree of freedom for each of the $N$ species, but a degree of freedom must be lost due to the sum constraint. As a result, the dimension of the niche space is, at most, $M \leq N -1$. From now on, we will assume that $M = N-1$ for simplicity; later, we will discuss how to remove irrelevant dimensions during data analysis. 

The Lagrange multipliers $\lambda_{\mu}$ encode the availabilities of the various resources. The values of the Lagrange multipliers have an interpretation that we can borrow from economics. A Lagrange multiplier is called a `shadow price' and is given by $\lambda_{\mu} = \frac{\del H}{\del V_{\mu}}$ evaluated at the optimum \cite{reznik2013flux}. That is, the shadow price describes the number of units that diversity changes if the availability of resource $\mu$ increases by one unit. The shadow price can be positive or negative. If the shadow price of resource $\mu$ is positive then an increase in the amount of resource $\mu$ will lead to an increase in the diversity of the community. By contrast, if the shadow price of resource $\mu$ is negative then an increase in the amount of resource $\mu$ will lead to a decrease in the diversity of the community. 

\subsection{Population dynamics that maximize diversity }

Alternatively, we can view Equation (\ref{eq:equil}) as the steady-state solution of an appropriately chosen set of dynamical equations. We stress that there are many types of dynamical models that lead to the same equilibrium, and our goal in this section is only to discuss the simplest dynamics that fulfill the principle of maximum diversity. 

First, we suppose that the fitness of a species can be represented as a weighted sum of its traits (i.e. resource utilizations) via $\mathcal{F}_i^* = \sum_{\mu} \lambda_{\mu} V_{i \mu}$. Then, we can write a simple equation for the dynamics of $\bvec{y}= \matr{G} \log \bvec{x}$ as: 
\be
\frac{ d}{dt} \bvec{y}  = \matr{G}\bvec{\mathcal{F}}^* - \bvec{y}
\label{eq:dyn}
\ee
where $\matr{G}$ is the matrix of the additive log-ratio transform. This equation reaches equilibrium at $\bvec{y}^* = \matr{G} \bvec{\mathcal{F}}^* = \matr{G} \log \bvec{x}^*$ and, thus, leads to the same equilibrium abundances as the principle of maximum diversity. This equation can also be written as $\log \bvec{x}^* = \bvec{\mathcal{F}}^* + \text{constant}$. Left multiplication by the non-invertible matrix $\matr{G}$ sends any constant term to zero, thereby removing any dependence on the total population size.

The equations for the population dynamics can be written in multiple forms that highlight different aspects of the model. For example, Equation (\ref{eq:dyn}) is equivalent to:
\be
\frac{ d}{dt} \bvec{y} = - \matr{G} \nabla_{\bvec{x}} D_{\text{KL}}(\bvec{x} || \bvec{x}^*) 
\ee
so that the community follows the gradient of the Kullback-Leibler divergence from its equilibrium configuration \cite{kullback_information_1951}. Similarly, the time derivatives of the relative abundances (rather than the additive log-ratio transformed abundances) can be obtained using the chain rule, and are given by a replicator equation of the form \cite{nowak2006evolutionary}:
\be
\frac{d}{dt} x_i = x_i  ( (\tilde{\mathcal{F}}_i^* - \tilde{\mathcal{F}}_i ) -  \sum_j x_j (\tilde{\mathcal{F}}_j^* - \tilde{\mathcal{F}}_j ))
\ee
where $\tilde{\bvec{\mathcal{F}}} = \matr{C} \log \bvec{x}$ is the vector of centered log-ratio transformed relative abundances.

\section{Modeling Variability in the Human Microbiota}

The previous section presented a theoretical model for determining the relative abundances of the species in a community based on a principle of maximum diversity. The principle postulates that the equilibrium relative species abundances in a community maximize the diversity of the community while ensuring that all effective resources are fully utilized. This section will explore the consequences of this model for variability in the composition of the human microbiota across different body sites, and across the population. 

The principle of maximum diversity implies that the additive log-ratio transform of the relative species abundances in a community is given by:
\be
\bvec{y} := \matr{G} \log \bvec{x} = \tilde{\matr{V}} \bvec{\lambda}
\ee
The $(N-1) \times (N-1)$ matrix $\tilde{\matr{V}} := \matr{G} \matr{V}$ describes intrinsic characteristics of the species, and does not vary from one environment to another. By contrast, the vector $\bvec{\lambda}$ encodes the information about the availabilities of the effective resources in a given environment. Thus, $\bvec{\lambda}$ varies from person-to-person, as well as across bodysites within the same individual. 

Variation from person-to-person of the composition of the microbiota in a single bodysite (such as the gut) is generally much smaller than the differences between bodysites within a single individual. Therefore, we propose a simple statistical model in which $\bvec{\lambda}$ is a random variable drawn from a distribution that depends on the bodysite. We hypothesize that, conditioned on the bodysite $s$, $\bvec{\lambda}$ is normally distributed via $\bvec{\lambda} | s \sim \mathcal{N}(\bar{\bvec{\lambda}}_s , \matr{\Sigma}_s)$ where $\matr{\Sigma}_s$ is a diagonal covariance matrix. For example, if we could determine the $\bvec{\lambda}$'s for the gut microbiomes of a large sample of people, we would find that $\lambda_{\mu}$ has a variance $\Sigma_{\mu \mu | s} = \sigma^2_{\mu |s }$ and that any two resources, say $\lambda_{\mu}$ and $\lambda_{\nu}$, are uncorrelated. Moreover, the additive log-ratio (ALR) transformed relative species abundances taken from that bodysite (e.g. the gut) would be normally distributed according to:
\be
\bvec{y} | s \sim \mathcal{N} (\tilde{\matr{V}} \bar{\bvec{\lambda}}_s , \tilde{\matr{V}} \matr{\Sigma}_s \tilde{\matr{V}}^T)
\ee
Thus, the covariance matrix ($\matr{\Psi}_s$) computed from the ALR transformed relative abundances is given by $\matr{\Psi}_s = \tilde{\matr{V}} \matr{\Sigma}_s \tilde{\matr{V}}^T$ for each body site $s$. 

The total covariance matrix of the ALR transformed abundances (i.e. the covariance matrix computed without using the bodysite labels) corresponds to:
\begin{align}
\text{Cov}[\bvec{y}] &:= \matr{\Psi} = \tilde{\matr{V}} (\matr{\Lambda} + \bar{\matr{\Sigma}}) \tilde{\matr{V}}^T
\end{align}
where 
\be
\matr{\Lambda}_{\mu \nu} := \text{Cov}[\bar{\lambda}_{\mu|s}, \bar{\lambda}_{\nu|s} ] = \sum_s p_s (\bar{\lambda}_{\mu |s} - \bar{\lambda}_{\mu} )( \bar{\lambda}_{\nu |s} - \bar{\lambda}_{\nu} )
\ee
with $\bar{\lambda}_{\mu} = \sum_s p_s \lambda_{\mu |s}$, and 
\be
\bar{\matr{\Sigma}}_{\mu \nu} := \text{E}[\matr{\Sigma}_{\mu \nu | s}] = \sum_s p_s \matr{\Sigma}_{\mu \nu | s}
\ee
where $p_s$ is the fraction of samples corresponding to bodysite $s$. In other words, $\matr{\Lambda}$ describes differences in the availabilities of the effective resources between different bodysites, while $\bar{\matr{\Sigma}}$ describes the variability in the availabilities of the resources within the bodysites. We hypothesize that $\bar{\matr{\Sigma}}$ is approximately diagonal, whereas $\matr{\Lambda}$ is likely not diagonal. As a result, it is possible to learn about the species characteristics by finding a matrix $\tilde{\matr{V}}$ such that $\tilde{\matr{V}}^{-1} \matr{\Psi}_s (\tilde{\matr{V}}^{T})^{-1}$ is approximately diagonal for each bodysite $s$. This trick does not tell us about $\matr{\Lambda}$, however, which has too many degrees of freedom to be determined from the data. Therefore, we expect that information about species characteristics (i.e. about $\tilde{\matr{V}}$) can be obtained by analyzing intra-bodysite variability in species composition, but not inter-bodysite variability.

Principal components analysis (PCA) is commonly applied to metagenomic data as a tool for uncovering an underlying structure, such as clustering of particular environments (e.g., \cite{arumugam2011enterotypes}). PCA decomposes the covariance matrix of ALR transformed relative abundances via $\matr{\Psi} = \tilde{\matr{V}}_{PCA} \matr{\Sigma}_{PCA} \tilde{\matr{V}}_{PCA}^T$, where $\matr{\Sigma}_{PCA}$ is diagonal and $\tilde{\matr{V}}_{PCA} \tilde{\matr{V}}_{PCA}^T = \matr{I}$. From Equation 10, we see that PCA is only relevant as a generative model if $\matr{\Lambda}$ is diagonal and $\tilde{\matr{V}}$ is orthogonal. Therefore, applying PCA to the covariance matrix computed without regard to the bodysites generally misses the structure that can be found by focusing only on intra-bodysite variability. In the next section, we describe a type of `generalized PCA' called Common Components Analysis (CCA) that uses the relationship $\matr{\Psi}_s = \tilde{\matr{V}} \matr{\Sigma}_s \tilde{\matr{V}}^T$ to infer the species characteristics from metagenomic data with labeled bodysites. 

\section{Common Components Analysis}

The model that we have presented makes it possible to use metagenomic data collected from a large sample of different individuals and bodysites to infer the matrix of resource utilizations ($\matr{V}$) and the Lagrange multipliers (i.e., shadow prices) ($\bvec{\lambda}$) that reflect the availabilities of the resources. The main assumptions are: (1) variation in species composition between samples (i.e., in different individuals and/or bodysites) is driven by variation in the availabilities of the effective resources, (2) the habitat variation can be captured by treating the shadow prices as random variables with a distribution that depends on the bodysite, and (3) the resource utilization matrix $\matr{V}$ is sparse so that any particular species is unlikely to be able to utilize all of the different resources.

\subsection{Inference Algorithm}

Common components analysis (CCA) is an approach to \emph{simultaneous non-orthogonal approximate diagonalization} \cite{flury1987two,vollgraf2006quadratic,trendafilov2010stepwise}. We have assumed that the distribution of ALR transformed species abundances is conditioned on the bodysite as $\bvec{y} | s = \tilde{\matr{V}} \bvec{\lambda}_s$ with $\bvec{\lambda}_s \sim \mathcal{N}( \bar{\bvec{\lambda}}_s, \matr{\Sigma}_s)$. Moreover, we assume that the bodysite labels are known. As in PCA, we make the assumption that $\matr{\Sigma}_s$ is diagonal, but we do not assume that $\tilde{\matr{V}}$ is orthogonal. This formulation of CCA aims to find a single, common set of factors that explains the variation within each bodysite $s = 1, \ldots, S$.  The factors are such that the product $\tilde{\matr{V}}^{-1} \matr{\Psi}_s (\tilde{\matr{V}}^{-1})^{T}$ is approximately diagonal for each $s$. 

Within a given bodysite, we have $\bvec{y} | s \sim \mathcal{N}(\tilde{\matr{V}} \bar{\bvec{\lambda}}_s | \eta \matr{I} + \tilde{\matr{V}} \matr{\Sigma}_s \tilde{\matr{V}}^T)$. Here, $\eta$ is a small noise term that accounts for experimental errors. In the following, we assume that the experimental errors are small relative to the intra-bodysite variation in the relative abundances so that they can be neglected, but we have included the term here for completeness. Maximizing the log-likelihood is equivalent to minimizing the KL-divergence between the assumed distribution and the empirical distribution. The distribution of $\bvec{y}$ given $s$ is multivariate normal, so the KL-divergence is given by:
\begin{widetext}
\begin{align}
\mathcal{L}_s( \bar{\bvec{\lambda}}_s, \tilde{\matr{V}}, \matr{\Sigma}_s ) &\propto \text{Tr}[ \matr{\Psi}_s ( \tilde{\matr{V}} \matr{\Sigma}_s \tilde{\matr{V}}^T )^{-1} ] + ( \bar{\bvec{y}}_s - \tilde{\matr{V}} \bar{\bvec{\lambda}}_s)^T ( \tilde{\matr{V}} \matr{\Sigma}_s \tilde{\matr{V}}^T )^{-1} ( \bar{\bvec{y}}_s - \tilde{\matr{V}} \bar{\bvec{\lambda}}_s) + \log | ( \tilde{\matr{V}} \matr{\Sigma}_s \tilde{\matr{V}}^T ) | - \log | \matr{\Psi}_s |
\end{align}
\end{widetext}
Here, $\bar{\bvec{y}}_s$ and $\matr{\Psi}_s$ are the empirical mean and covariance matrix of $\bvec{y} = \matr{G} \log \bvec{x}$ in bodysite $s$, respectively. We can immediately see that $\bar{\bvec{\lambda}}_s = \tilde{\matr{V}}^{-1} \bar{\bvec{y}}_s$. Plugging this in, rearranging, and neglecting constant terms gives, 
\begin{align}
 \mathcal{L}_s(\tilde{\matr{V}}, \matr{\Sigma}_s ) \propto \text{Tr}[ \matr{\Psi}_s ( \tilde{\matr{V}} \matr{\Sigma}_s \tilde{\matr{V}}^T )^{-1} ] - \log | ( \tilde{\matr{V}} \matr{\Sigma}_s \tilde{\matr{V}}^T )^{-1} |
 \end{align}
which is the appropriate negative log-likelihood for a single bodysite.

The fraction of samples coming from bodysite $s$ is $p_s$, and the bodysite labels are known. Therefore, the total negative log-likelihood is a weighted sum of each of the individual negative log-likelihoods. The matrices $\tilde{\matr{V}}$ and $\{ \matr{\Sigma}_s\}_{s=1}^S$ can be inferred by minimizing this conditional negative log-likelihood:
\begin{align}
 \mathcal{L}(\tilde{\matr{V}}, \{ \matr{\Sigma}_s\}_{s=1}^S ) &= \sum_s p_s (\text{Tr}[ \matr{\Psi}_s ( \tilde{\matr{V}} \matr{\Sigma}_s \tilde{\matr{V}}^T )^{-1} ] \nonumber \\
 &- \log | ( \tilde{\matr{V}} \matr{\Sigma}_s \tilde{\matr{V}}^T )^{-1} |)
 \end{align}
We can simplify this by redefining the objective function in terms of $\matr{W}^T = \tilde{\matr{V}}^{-1}$, giving:
\begin{align}
\mathcal{L}(\matr{W}, \{ \matr{\Sigma}_s^{-1} \}_{s=1}^S ) &= \sum_s p_s (\text{Tr}[ \matr{\Psi}_s \matr{W} \matr{\Sigma}_s^{-1} \matr{W}^{T} ] \nonumber \\
&- \log | \matr{W} \matr{\Sigma}_s^{-1} \matr{W}^{T} |)
\end{align}
The derivatives of the objective function can be calculated as:
\begin{align}
&\frac{ \del \mathcal{L} } { \del \matr{W} } = 2 \sum_s p_s (\matr{\Psi}_s \matr{W} \matr{\Sigma}_s^{-1} - (\matr{W}^{-1})^T)  \equiv \Delta \matr{W} \\
&\frac{ \del \mathcal{L} } { \del \matr{\Sigma}_s^{-1} } = \text{diag} ( \matr{W}^T \matr{\Psi}_s \matr{W} - \matr{\Sigma}_s ) \equiv \Delta \matr{\Sigma}^{-1}_s
\end{align}
The objective function can be minimized by alternating gradient descent updates $\matr{W}(t+1) = \text{norm}(\matr{W}(t)  - \e_w ( \Delta \matr{W}(t)  + \rho \Delta \matr{W}(t-1) ))$ and $\matr{\Sigma}_s^{-1} (t+1) = \matr{\Sigma}_s^{-1}(t) - \e_s (\Delta \matr{\Sigma}_s^{-1} (t) + \rho \Delta \matr{\Sigma}_s^{-1} (t-1) )$ where $\rho \approx 0.95$ is a momentum term and $\text{norm}(\cdot)$ normalizes the columns of the matrix \cite{qian1999momentum}. Normalizing the columns of $\matr{W}$ is an arbitrary choice that resolves an inherent ambiguity in which it is not possible to determine the relative magnitudes of $\tilde{\matr{V}}$ and $\matr{\Sigma}_s$.  Alternating the updates of $\matr{W}$ and $\matr{\Sigma}_s^{-1}$ makes it easier to tune the step sizes $\e_w$ and $\e_s$ using the `bold driver' method \cite{vogl1988accelerating} where we only accept gradient steps that decrease the objective function, and we increase the step size (e.g. $\e_w \leftarrow 1.1*\e_w$) if a step in in the $\matr{W}$ direction is accepted and decrease the step size (e.g. $\e_w \leftarrow 0.5*\e_w$) if a step in the $\matr{W}$ direction is rejected, and similarly for $\e_s$. The gradient descent is continued until convergence.

\subsection{Sparse Recovery}

The CCA algorithm described above yields an estimate for $\tilde{\matr{V}} = (\matr{W}^T)^{-1} = \matr{G} \matr{V}$. We would like to be able to determine $\matr{V}$ but the matrix $\matr{G}$ is not invertible. However, if we assume that $\matr{V}$ is sparse then it is possible to determine $\matr{V}$ from $\tilde{\matr{V}}$. In the context of the model, this assumption means that any individual species is unlikely to be able to utilize every possible resource. To recover $\matr{V}$ from $\tilde{\matr{V}}$, we solve the problem:
\be
\min{ || \matr{V} ||_1} \text{ subject to } \matr{G} \matr{V} = \tilde{\matr{V}}
\ee
where $||\matr{V} ||_1 = \sum_{i\mu} |V_{i\mu} |$. The solutions to this problem are all of the form $V_{i \mu} = z_{\mu} + \tilde{V}_{(i -1), \mu} (1 - \d_{i 1})$ for $i = 1, \ldots, N$, where $z_{\mu}$ can, in principle, take on any real value. Because we want the solution with a minimum $L_1$ norm, it is sufficient to test $z_{\mu} = 0$ and $z_{\mu} \in \{ -\tilde{V}_{i, \mu} \}_{i = 1}^{N-1}$ (the only sparse solutions) and to choose the one with minimum norm. This is a tractable search over $N(N-1)$ possibilities in the worst case and can be done easily for reasonable system sizes. 

\subsection{Number of Components}

We have assumed that the matrix $\tilde{\matr{V}}$ has dimension $(N-1) \times (N-1)$ and is full-rank or, equivalently, that the number of effective resources is $M = N-1$. This assumption could be relaxed by including the term accounting for experimental errors (i.e., setting $\eta > 0$). Even without the full error model, however, one may find that many of the components have low weight. That is, there may be many $\mu$'s with small contributions to the intra-bodysite variances (e.g., small $\sum_s p_s \matr{\Sigma}_{\mu \mu |s}$). This is similar to the usual implementation of PCA where many eigenvectors of the covariance matrix may be associated with small eigenvalues. The lowly weighted components are often neglected to obtain a low-rank approximation of the data. 

In the case of CCA, however, choosing which components to keep depends on the objective. For example, to describe the variation within bodysite $s$ we would want to keep the components with the largest values of $\matr{\Sigma}_{\mu \mu | s}$ because we don't care about the other bodysites. By contrast, components can be chosen to obtain a low dimensional representation that clusters the bodysites by choosing components with large inter-bodysite differences and small intra-bodysite variances. One way to choose these components is to sort them by $\sum_s p_s (\bar{\lambda}_{\mu | s} - \bar{\lambda}_{\mu})^2 / \sum_s p_s \Sigma_{\mu \mu | s}$ and keep only the components with the largest scores. This is the technique used to choose the components in Figure 2e of the main text. 

\subsection{Implementation}

Code for Common Components Analysis was written in Python (version 3.43) and uses NumPy. The source code is available at https://sites.google.com/site/charleskennethfisher/.   

\begin{figure}
\centering
\includegraphics[width=\linewidth]{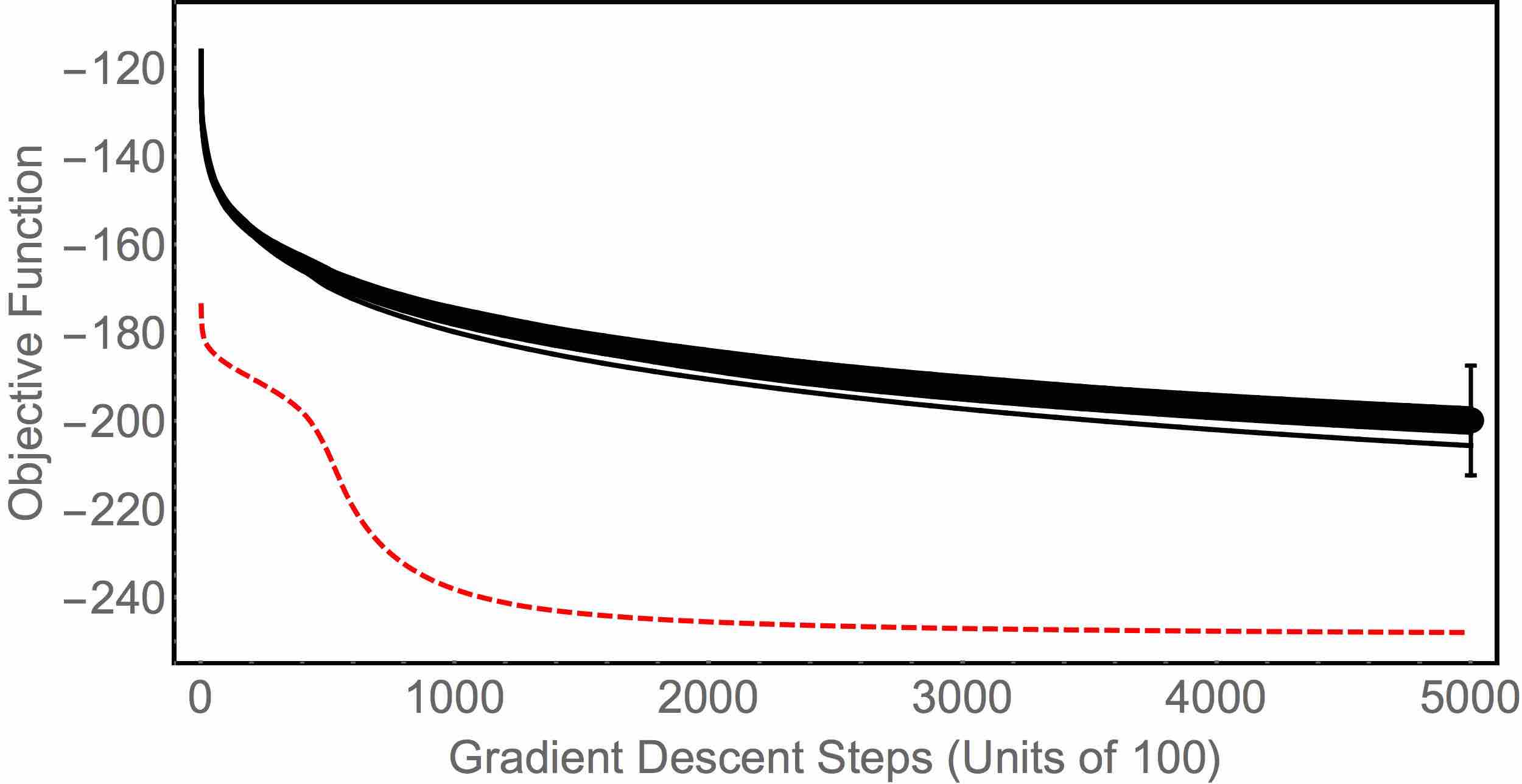}
\caption{ \textbf{Fitting the common components analysis model.} Plot of the CCA objective function during gradient descent using the true covariance matrices (red, dashed line) and 20 randomized covariances matrices (black lines). The error bars on the final value of the objective function with the randomized matrices represent $\pm$ 6 standard deviations. }
\label{fig:figS1}
\end{figure}

\section{Data used in this study}

We analyzed data on relative species abundances collected as part of the Human Microbiome Project (HMP) \cite{turnbaugh2007human,human2012structure,human2012framework,gevers2012human}. Data on species composition from the HMP were downloaded from the MG-RAST server (project 385) \cite{meyer2008metagenomics,glass2010using}. These metagenomic data correspond to 1606 human microbiota samples separated into four main body sites: gut, oral, skin, and vaginal. All unclassified and non-bacterial species were removed from the data. Each species ($i$) was assigned a rank ($r_i^l$) in each bodysite ($l$) based on decreasing relative abundance; e.g. the most abundant species in sample $l$ was assigned $r_i^l =1$. Then, the median rank for each species was computed over all 1606 samples. The 100 species with the smallest median ranks (i.e., the 100 most abundant species) were selected for further study in order to make computational studies more feasible. Selecting species in this manner (i.e., by typical rank) ensures that every species is present in each bodysite. After performing the species selection procedure, two of the oral microbiota samples (MG-RAST ID 4472526.3 and 4472527.3) did not have enough species counts and were discarded. Thus, we analyzed data on the relative abundances of 100 species in 1604 samples taken from 4 different bodysites. Species compositions were computed using a pseudocount of $0.5$ via:
\be
x_i = \frac{ n_i + 0.5} {\sum_j (n_j + 0.5)}
\ee
to ensure that $x_i > 0$, which is necessary for taking the logarithm as part of the ALR transform. The raw species counts used in this study are available in a text file at https://sites.google.com/site/charleskennethfisher/.

\section{Summary: Inference and Model Validation}

\textbf{Inference:}
\begin{enumerate} 
\item Compute the emprical covariance matrices $\matr{\Psi}_s = \text{Cov}[ \matr{G} \log \bvec{x} | s]$ for each bodysite $s$. 
\item Use CCA to identify matrices $\tilde{\matr{V}}$ and $\{ \matr{\Sigma}_s \}_{s=1}^S$ so that $\matr{\Psi}_s \approx \tilde{\matr{V}} \matr{\Sigma}_s \tilde{\matr{V}}^T$. 
\item Solve the sparse recovery problem $\matr{G} \matr{V} = \tilde{\matr{V}}$ for $\matr{V}$. 
\end{enumerate}
\textbf{Validation:}
\begin{enumerate} 
\item Check goodness-of-fit $\matr{\Psi}_s \approx \tilde{\matr{V}} \matr{\Sigma}_s \tilde{\matr{V}}^T$.
\item Check that $\text{Var}[\lambda_{\mu} | s] \approx \sigma^2_{\mu|s}$.
\item Check that $\text{Corr}[ \lambda_{\mu}, \lambda_{\nu} | s] \approx 0$.
\item Check that species similarities computed from $\matr{V}$ are correlated to taxonomic similarities.
\end{enumerate}
\textbf{Randomization:}
\begin{enumerate} 
\item Generate randomized covariance matrices $\matr{\Psi}_{s}^{'}$ (see next section). 
\item Use CCA to identify matrices $\tilde{\matr{V}}$ and $\{ \matr{\Sigma}_s \}_{s=1}^S$ so that $\matr{\Psi}_{s}^{'} \approx \tilde{\matr{V}} \matr{\Sigma}_s \tilde{\matr{V}}^T$. 
\item Solve the sparse recovery problem $\matr{G} \matr{V} = \tilde{\matr{V}}$ for $\matr{V}$. 
\item Verify that the CCA model validation steps fail with randomized covariance matrices. 
\end{enumerate}

\section{Measuring Goodness-of-fit for CCA}

\begin{figure*}
\centering
\includegraphics[width=\textwidth]{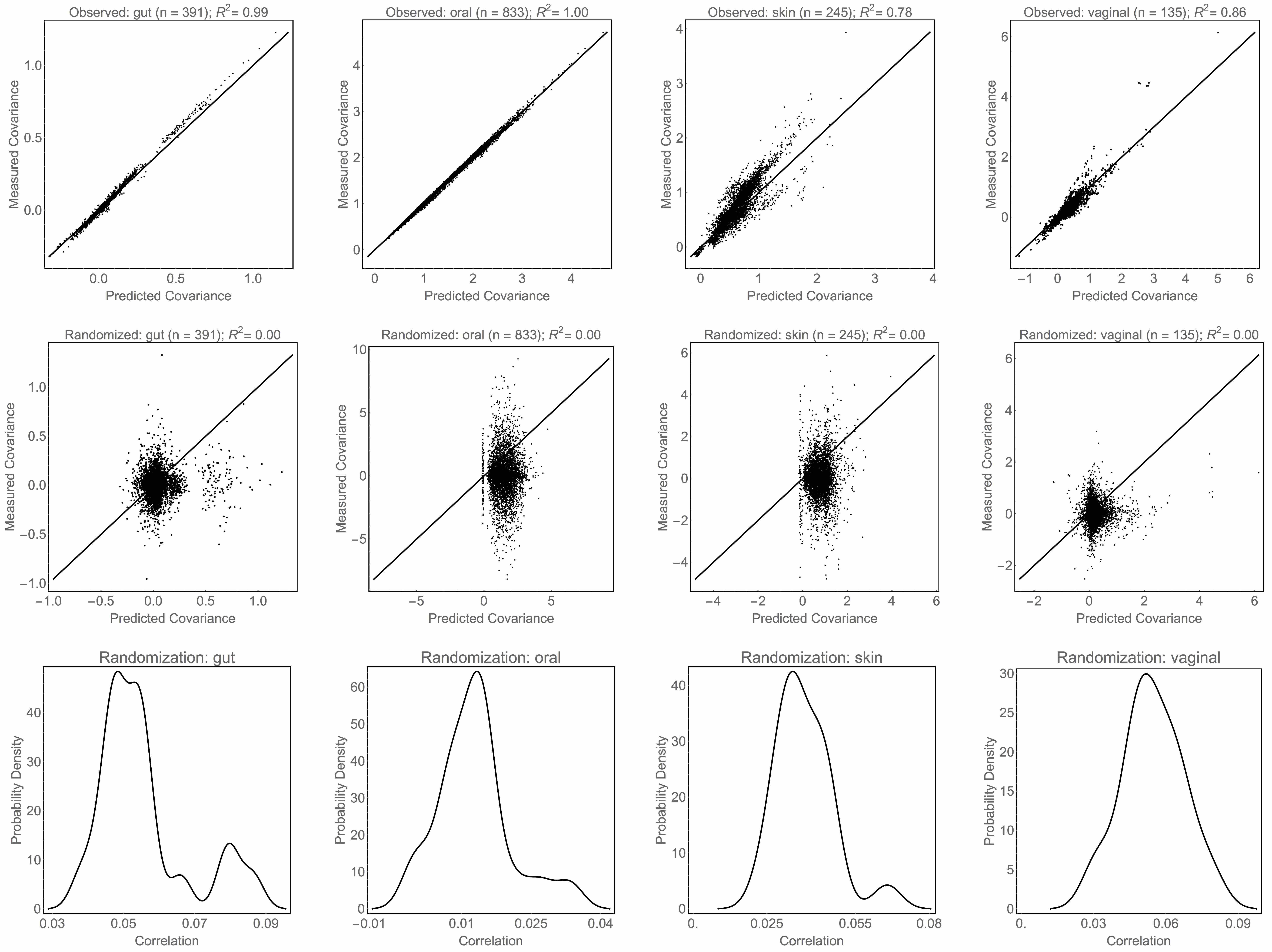}
\caption{ \textbf{Goodness-of-fit of the common components analysis model.}  (Top row) Correlations between the observed covariances ($\matr{\Sigma}_s$) and those computed from CCA ($\tilde{\matr{V}} \matr{\Sigma}_s \tilde{\matr{V}}^T$) in each of the 4 body sites. (Middle row) Correlations between the randomized covariances ($\matr{\Sigma}_s$) and those computed from CCA ($\tilde{\matr{V}} \matr{\Sigma}_s \tilde{\matr{V}}^T$) in each of the 4 body sites. Shows only the best fit of the 20 randomizations. (Bottom row) The  distribution of correlations from all 20 randomizations, which are roughly an order of magnitude smaller than the observed values.  }
\label{fig:figS2}
\end{figure*}

\subsection{Strategy for randomization}

Validating the model underlying CCA requires a test that the covariances matrices ($\matr{\Psi}_s$) are, indeed, approximately simultaneously diagonalizable. This requires two separate tests. First, we need to see if it is even possible to find an $(N-1) \times (N-1)$ matrix $\tilde{\matr{V}}$ and diagonal matrices $\matr{\Sigma}_s$ such that $\matr{\Psi}_s = \tilde{\matr{V}} \matr{\Sigma}_s \tilde{\matr{V}}^T$ for each body site $s \in \{ \text{gut}, \text{oral}, \text{skin}, \text{vaginal} \}$. Second, we need show that it is usually not possible to simultaneously diagonalize 4 randomly generated covariance matrices. This will establish that simultaneous diagonalizability is a special property of the observed covariance matrices.  

In order to test the second point, we ran the CCA algorithm multiple times using randomly generated covariance matrices. These randomized covariances matrices were constructed as follows. Let $\matr{\Sigma}_{PCA}^{(s)}$ denote the matrix with the eigenvalues of $\matr{\Psi}_s$ along the diagonal. The randomized covariance matrix is given by $\matr{\Psi}_s^{'} = \matr{Q} \matr{\Sigma}_{PCA}^{(s)} \matr{Q}^T$ where $\matr{Q}$ is a random orthogonal matrix \cite{genz1998methods}. This randomization procedure ensures that all of the random covariance matrices for a given bodysite have the same eigenvalues, but different eigenvectors. This is the simplest null model for testing simultaneous diagonalizability. 

Due to computational expense of CCA (at least, given the current implementation), it was not possible to perform the thousands of iterations that would be necessary to compute accurate p-values. Therefore, we ran the CCA algorithm on 20 random realizations of the covariance matrices and report the mean and standard deviation, or histograms, for each of the quantities that we computed from the observed data, as noted in the figure legends. 

\subsection{Results}

\begin{figure*}
\centering
\includegraphics[width=\textwidth]{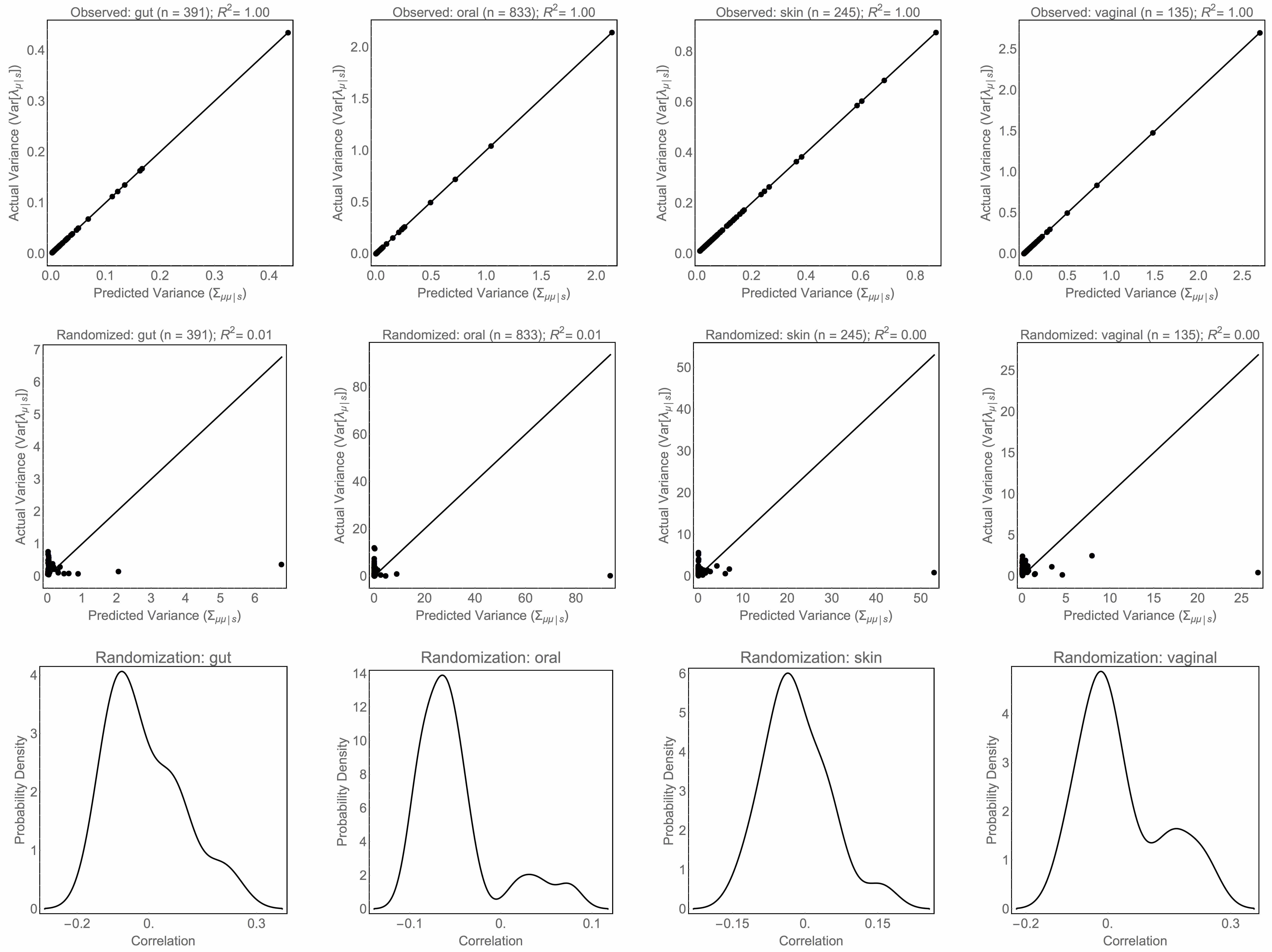}
\caption{ \textbf{Comparing the variance in $\bvec{\lambda}$ to $\matr{\Sigma}_s$.}  (Top row) Correlations between diagonal elements of $\matr{\Sigma}_s$ and the variances computed from the inferred $\bvec{\lambda}$'s. (Middle row) Correlations between diagonal elements of $\matr{\Sigma}_s$ and the variances computed from the inferred $\bvec{\lambda}$'s for the best of the 20 randomizations. (Bottom row) The distribution of correlations from all 20 randomizations. }
\label{fig:figS3}
\end{figure*}

\begin{figure*}
\centering
\includegraphics[width=\textwidth]{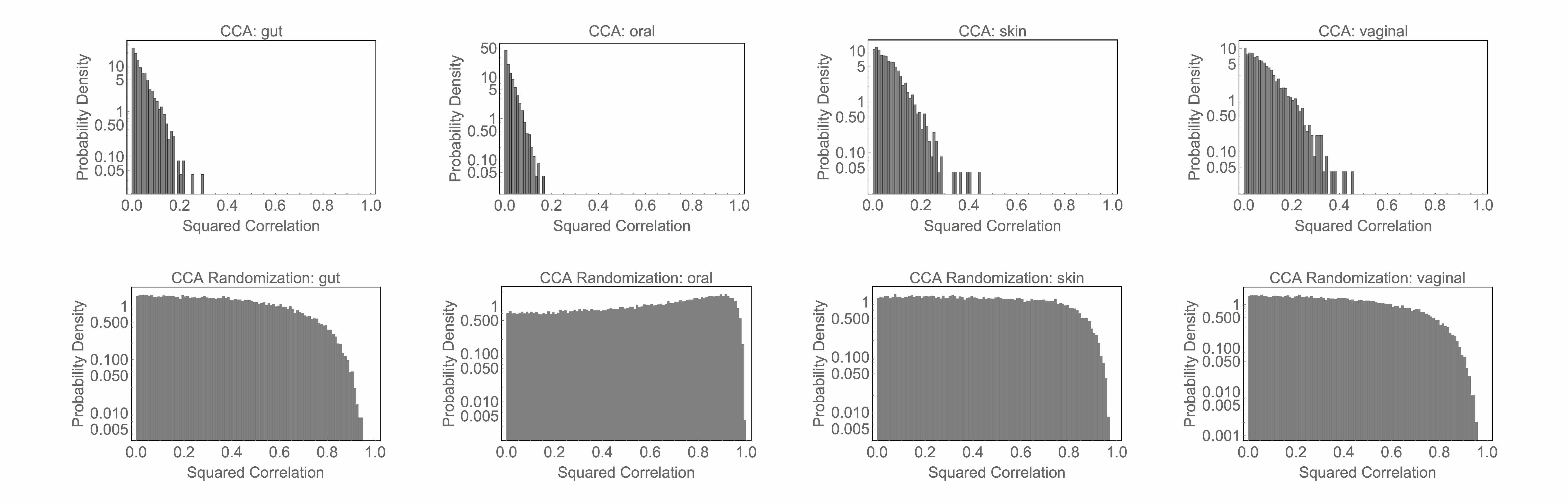}
\caption{ \textbf{Intra-body site correlations between common components.} Histograms of the correlations between $\lambda_{\mu}$ and $\lambda_{\nu}$, conditioned on body site, computed from observed covariance matrices (top row) and randomized covariance matrices (bottom row). These plots show that the niche availabilities obtained from the observed data are approximately uncorrelated, whereas those inferred from randomized covariance matrices are not.  }
\label{fig:figS4}
\end{figure*}

A plot of the objective function during gradient descent is shown in Figure \ref{fig:figS1}. The objective function computed from the observed covariance matrices converges to a value many standard deviations below the objective functions obtained with randomized covariance matrices. The best fit parameters from CCA reproduce the observed covariances with reasonable accuracy, as illustrated by the Pearson correlations of $R^2 = 0.99$ for the gut samples, $R^2 = 1.00$ for the oral samples, $R^2 = 0.78$ for the skin samples, and $R^2 = 0.86$ for the vaginal samples (see top row of Figure \ref{fig:figS2}). Note that the body sites with larger sample sizes ($N_{gut} = 391$, $N_{oral} = 833$, $N_{skin} = 245$, $N_{vaginal} = 135$) have better fits. Moreover, the correlations between the observed and predicted covariances are many standard deviations larger than those obtained from the randomizations (see middle and bottom rows of Figure \ref{fig:figS2}). These results demonstrate that the observed covariances are, indeed, approximately simultaneously diagonalizable. 

Once $\tilde{\matr{V}}$ had been inferred, the shadow prices can be estimated as $\bvec{\lambda} = \tilde{\matr{V}}^{-1} \bvec{y}$. If the CCA model is correct, then the shadow prices from a particular body site $s$ should be uncorrelated and should have covariance matrix $\matr{\Sigma}_s$. The top row of Figure \ref{fig:figS3} shows a comparison between $\text{Var}[\lambda_{\mu} | s]$ and $\matr{\Sigma}_{\mu \mu |s}$. The Pearson correlations between the observed and predicted values are equal $R^2 = 1.00$ for each of the body sites. However, the middle and bottom rows of Figure \ref{fig:figS3} show that this relationship breaks down for the randomized covariance matrices because they are not simultaneously diagonalizable. 

As a final check to ensure that the CCA model is appropriate for the HMP data, we computed histograms of the correlations between shadow prices (i.e., $\text{Corr}[ \lambda_{\mu}, \lambda_{\nu} | s]$) within each body site. Ideally, all of the correlations would be zero but, in practice, there are weak correlations between some of the niches, as shown in the top line of Figure \ref{fig:figS4}. Histograms of the shadow price correlations obtained from the randomization experiments are shown in the bottom line of Figure \ref{fig:figS4} for comparison. Clearly, the correlations from the observed shadow prices are very weak compared to those obtained via randomization. 

\begin{figure}
\centering
\includegraphics[width=\linewidth]{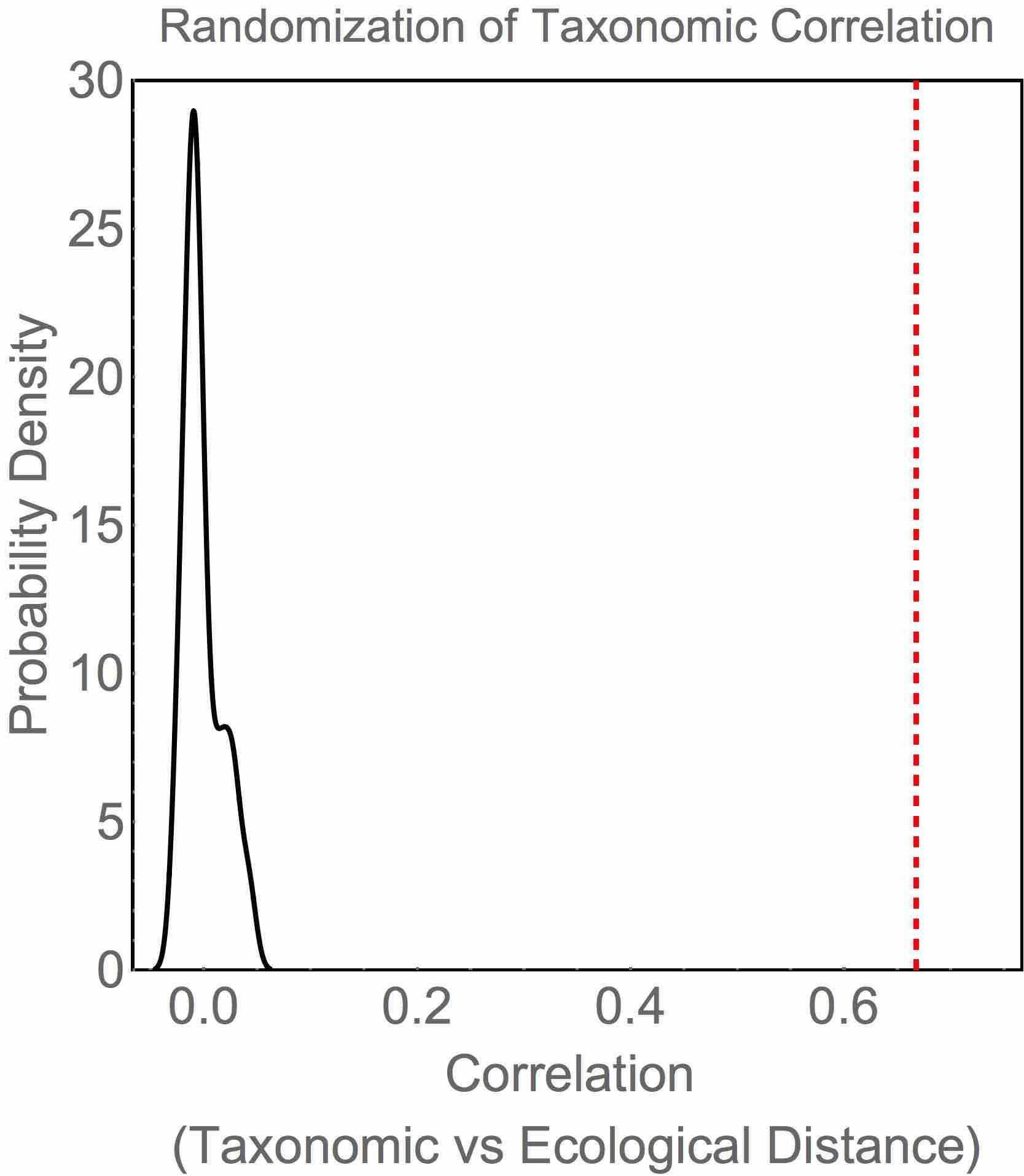}
\caption{ \textbf{Correlation between taxonomic and ecological distances.} Histogram of the correlation between the taxonomic distance and ecological distances computed from randomized covariance matrices. The correlation obtained with the observed data is shown as a dotted red line. The true correlation lies far outside the distribution obtained from randomization.}
\label{fig:figS5}
\end{figure}

The final claim that we test via randomization is that there is a significant correlation between the distance between species computed from their taxonomic classifications and the distance between species computed from their common components. The taxonomic distance between two species $i$ and $j$ was calculated as $D_{ij}^{tax} = 5 - O_{ij}$ where $O_{ij}$ is the number of taxonomic levels shared by species $i$ and $j$. For example, two species that belong to the same genus have taxonomic distance $D_{ij}^{tax} = 0$, whereas two species that belong to different phyla have taxonomic distance $D_{ij}^{tax} = 5$. The taxonomic classifications used in this study are included as a text file in the Supplementary Information. Each species can be represented as a vector in the CCA derived niche space $\bvec{v}_i = \{ \bar{\sigma}_{\mu} \matr{V}_{i \mu} \}_{\mu = 1}^{N-1}$ where $\bar{\sigma}_{\mu}^2 = \sum_s p_s \matr{\Sigma}_{\mu \mu | s}$. This vector describes the ability of a species to utilize each of the effective resources, weighted by the their average variabilities. The ecological distance between two species $i$ and $j$ was computed from their common components using the `correlation distance,' $D_{ij}^{CCA} = 1 - \text{Corr}[ \bvec{v}_i , \bvec{v}_j ] = 1 - \cos (\theta_{ij})$ where $\theta_{ij}$ is the angle between vectors $\bvec{v}_i$ and $\bvec{v}_j$. The correlation between the taxonomic distance and the ecological distance computed for the observed covariance matrixes is $R = 0.67$. For comparison, a histogram of the correlations obtained from the randomization experiments is shown in Figure \ref{fig:figS5}. The observed correlation is many standard deviations from the values obtained though randomization. 

\section{Comparison with Principal Components Analysis}

\begin{figure*}
\centering
\includegraphics[width=\textwidth]{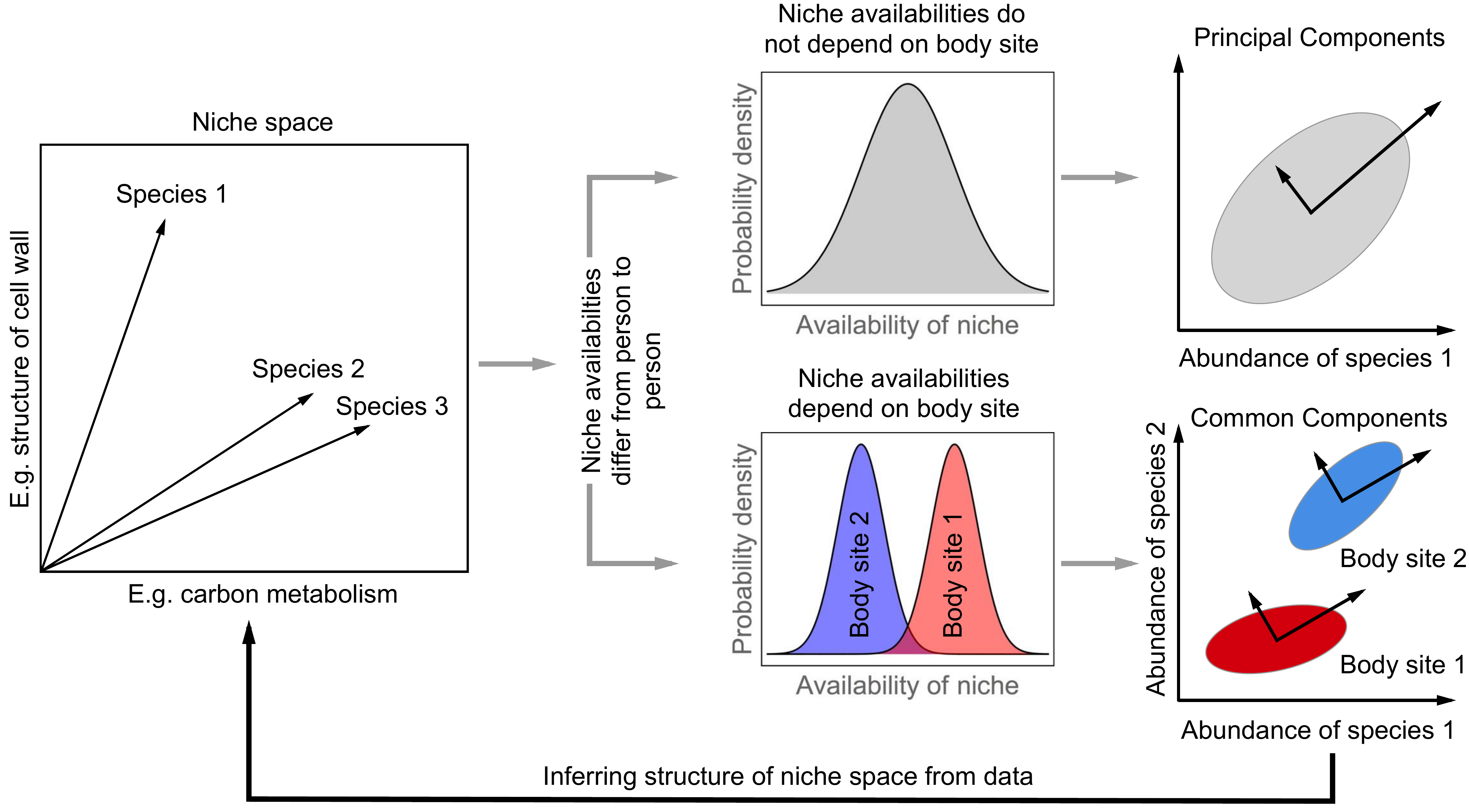}
\caption{\textbf{Schematic comparison of CCA and PCA.} Each species corresponds to a point in a high dimensional niche space. Fluctuations in the availabilities of the niches from person-to-person cause fluctuations in the relative abundances of the species. If the distribution of niche availabilities does not depend on the bodysite (e.g. gut, skin, etc) then the log-ratio transformed abundances are Gaussian distributed, and the structure of niche space can be inferred using Principal Components Analysis (PCA) by finding the set of axes with the largest variation. If the distribution of the niche availabilities does depend on the bodysite, however, then the log-ratio transformed abundances are drawn from mixture of Gaussians and maximum likelihood fitting of the model identifies a common set of axes, or common components, that approximately diagonalize the covariance matrices in each of the bodysites.}
\label{fig:figS6}
\end{figure*}

Principal Components Analysis (PCA) is a technique that is often applied to find low-dimensional representations of microbiota data. Common Components Analysis (CCA) shares some similarities with PCA so it is useful to compare the two techniques. The generative model for PCA is quite simple: the observed data $\bvec{y}$ arise from a Gaussian distributed latent variable $\bvec{\lambda}$ via $\bvec{y} = \tilde{\matr{V}} \bvec{\lambda}$. Here, the elements of, say $\lambda_{\mu}$ and $\lambda_{\nu}$, are uncorrelated and $\tilde{\matr{V}}$ is orthogonal. Thus, the main difference between the generative model for PCA and the generative model for CCA is that $\bvec{\lambda}$ is assumed to be Gaussian distributed in PCA, whereas $\bvec{\lambda}$ is drawn from a mixture of Gaussians in CCA (Figure \ref{fig:figS6}). As a result, PCA finds a set of directions that explain the total variation over all of the samples, whereas CCA finds a set of directions that explain the intra-bodysite variances.

\begin{figure*}
\centering
\includegraphics[width=\textwidth]{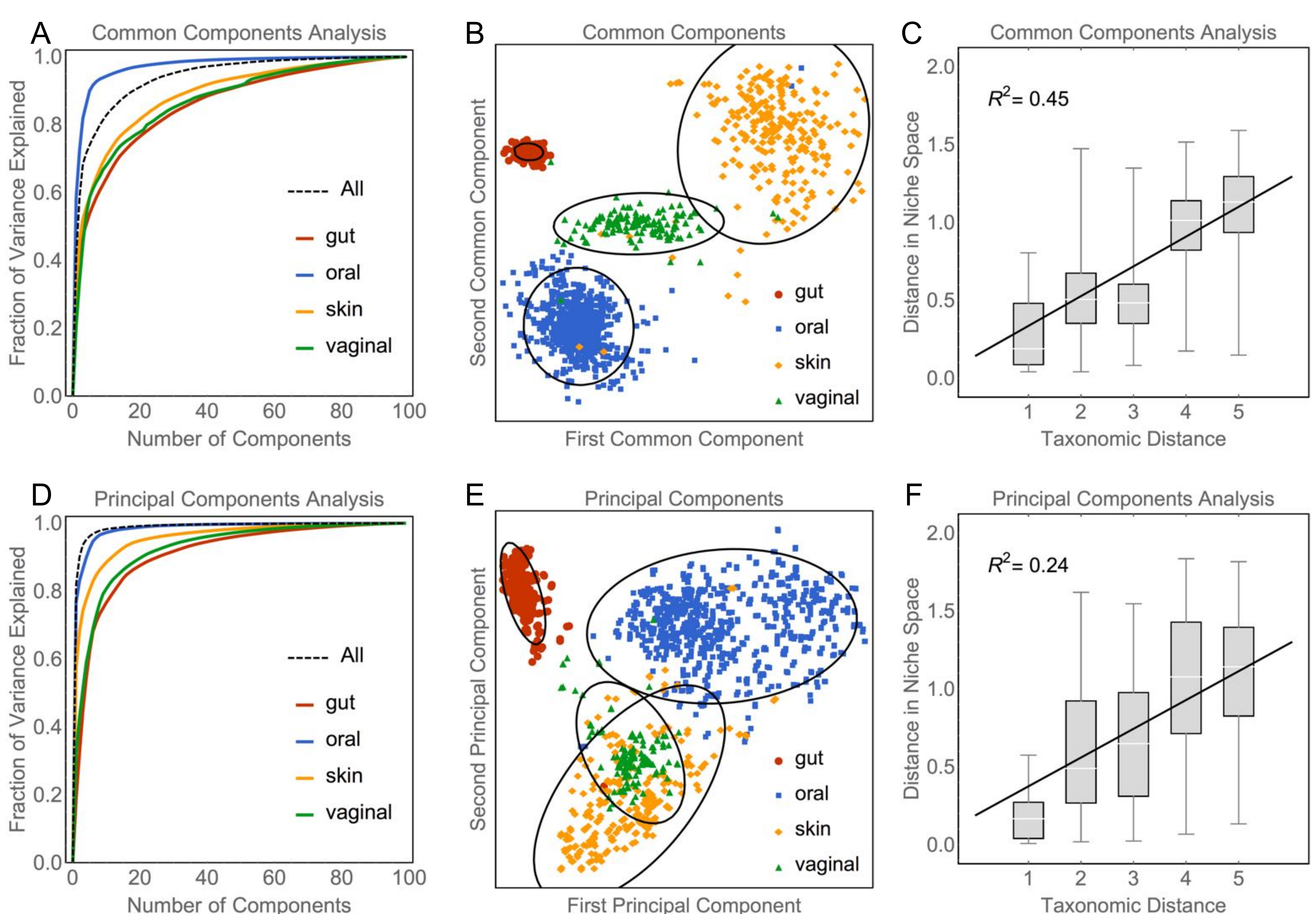}
\caption{ \textbf{Comparison of CCA and PCA on the HMP data.}  A) Percentage of variance explained in each bodysite as function of the number of common components. B) Projecting onto two common components with large inter-bodysite differences and small intra-bodysite variation separates the bodysites into coherent clusters. C) Distances between species computed from CCA are strongly correlated with taxonomy. Note that parts A-C are reproduced from the Main Text to facilitate comparison with PCA. D) Percentage of variance explained in each bodysite as function of the number of principal components. E) Projecting onto the two largest principal components fails to separate the bodysites into coherent clusters. F) Distances between species computed from PCA are only weakly correlated with taxonomy.}
\label{fig:figS7}
\end{figure*}

Figure \ref{fig:figS7} presents a comparison of PCA and CCA on the HMP data. The assumptions of the generative model for PCA are clearly violated by the HMP data; the bodysites form coherent clusters and, therefore, cannot be driven by Gaussian distributed resources. Nevertheless, PCA can still be used as a technique for dimensionality reduction and data visualization. Both PCA and CCA are able to explain compositional variation in the microbiota with a few components (Figure \ref{fig:figS7}a,d). However, the two largest principal components fail to separate all four bodysites (Figure \ref{fig:figS7}e). PCA fails to separate the bodysites because it finds directions that explain total variability, which is a sum of inter-bodysite differences and intra-bodysite variation. Thus, directions with the largest principal values may have both large inter-bodysite differences and large intra-bodysite variation. Separating the bodysites, however, requires one to identify directions with large inter-bodysite differences and small intra-bodysite variation. Thus, CCA is able to identify directions that cluster the bodysites more effectively.  

\section{Finding Significant Pathways with the Bayesian Ising Approximation}

\subsection{Obtaining Functional Pathways from KEGG}

The KEGG orthologies (Brite hierarchy 00001) for every available strain of each of the 100 species were downloaded from the KEGG database (http://www.genome.jp/kegg/kegg2.html, \cite{kanehisa2000kegg}). For each enzyme $\a$ (EC number) in the KEGG orthology and for each species $i$, we computed the fraction $f_{i \a}$ of strains that possessed that enzyme. For example, EC:1.1.1.1 (Alcohol dehydrogenase) was present in every strain of \emph{Streptococcus pneumoniae} giving it a value of $f_{i \a } = 1$. In this way, we constructed a matrix describing the presence/absence of each of the enzymes in all 100 species. 

Enzymes in the KEGG database are also organized into functional pathways such as Lysine biosynthesis and Glycolysis/Gluconeogenesis. For each pathway, we computed a distance between species as:
\be
D_{ij}^{\text{path}} = \sum_{\a \in \text{path}} ( f_{i \a} - f_{j \a})^2
\ee 
We performed a simple regression analysis to assess which of the KEGG pathways were associated with the ecological distance computed with CCA by analyzing the linear model:
\be
(D_{ij}^{CCA})^2 = \text{constant} +  \sum_{\text{path}} \beta_{\text{path}} (D_{ij}^{\text{path}})^2 + \eta_{ij}
\label{eq:regress}
\ee
where $\eta_{ij}$ is the residual noise. It is important to note that we make a simplifying assumption that the noise ($\eta_{ij}$) terms are identically and independently distributed normal random variables.

\subsection{Selecting Relevant Pathways with Bayesian Linear Regression}

We assessed the relevance of each pathway using a framework based on Bayesian statistics. The goal is to compute a `posterior' probability that the coefficient associated with a pathway is not equal to zero (i.e., $P(\beta_{\text{path}} \neq 0 | D^2_{CCA})$ where $D^2_{CCA}$ is the matrix of squared distances computed from CCA). However, because it is computationally challenging to compute exact posterior probabilities for variable selection we a use recently described approach called the Bayesian Ising Approximation (BIA) \cite{fisher2015bayesian,fisher2014bayesian}. For the sake of completeness, we provide a brief description of the BIA below -- more detailed results are described by Fisher and Mehta \cite{fisher2015bayesian}.

It will be helpful to introduce the half vectorization operator $\text{vech}(\text{mat})$ that takes the elements below the diagonal from each column in the matrix $\text{mat}$ and stacks them into a vector. Using this notation, we define $ \bvec{y} = \text{vech}(D^2_{CCA})$ and the design matrix $\matr{X}$ as the matrix with columns $\text{vech}(D^2_{\text{path}})$. Moreover, we standardize $\bvec{y}$ and each column of $\matr{X}$ to have zero mean and unit variance, which eliminates the need for the constant term in Equation \ref{eq:regress}.

Bayesian methods combine the information from the data, described by the likelihood function, with \emph{a priori} knowledge, described by a prior distribution, to construct a posterior distribution that describes one's knowledge about the parameters after observing the data. In the case of linear regression, the likelihood function is a Gaussian:
\be
P(\bold{y} | \bvec{\b}, \s^2) \propto \exp \left( - \frac{(\bold{y} - X \bvec{\b})^T (\bold{y} - X \bvec{\b})} {2\s^2} \right) \nonumber
\ee
In this work, we will use standard conjugate prior distributions for $\bvec{\b}$ and $\s^2$ given by $P(\bvec{\b},\s^2 | \bold{s}) = P(\s^2) P(\bvec{\b}|\s^2, \bvec{s})$ where:
\begin{align}
P(\s^2) &\propto (\s^2)^{-(a_0 + 1)} \exp(-b_0 / \s^2 ) \nonumber \\
P_{\lambda}(\bvec{\b} | \s^2, \bvec{s}) &\propto \prod_j \left[ (1-s_j) \d(\b_j ) + (1+s_j) \sqrt{ \frac{\l}{2 \pi \s^2}} \exp\left(-\frac{\l \b_j^2}{2 \s^2}\right) \right] \nonumber
\end{align}
Here, we have introduced a vector ($\bold{s}$) of indicator variables so that $\b_j = 0$ if $s_j = -1$ and $\b_j \neq 0$ if $s_j = +1$ for the $j^{th}$ pathway. We also have to specify a prior for the indicator variables, which we will set to a flat prior $P(\bold{s}) \propto 1$ for simplicity. In principle, $a_0$, $b_0$ and the penalty parameter on the regression coefficients, $\l$, are free parameters that must be specified ahead of time to reflect our prior knowledge. We will discuss these parameters in the next section. 

\begin{figure}
\centering
\includegraphics[width=\linewidth]{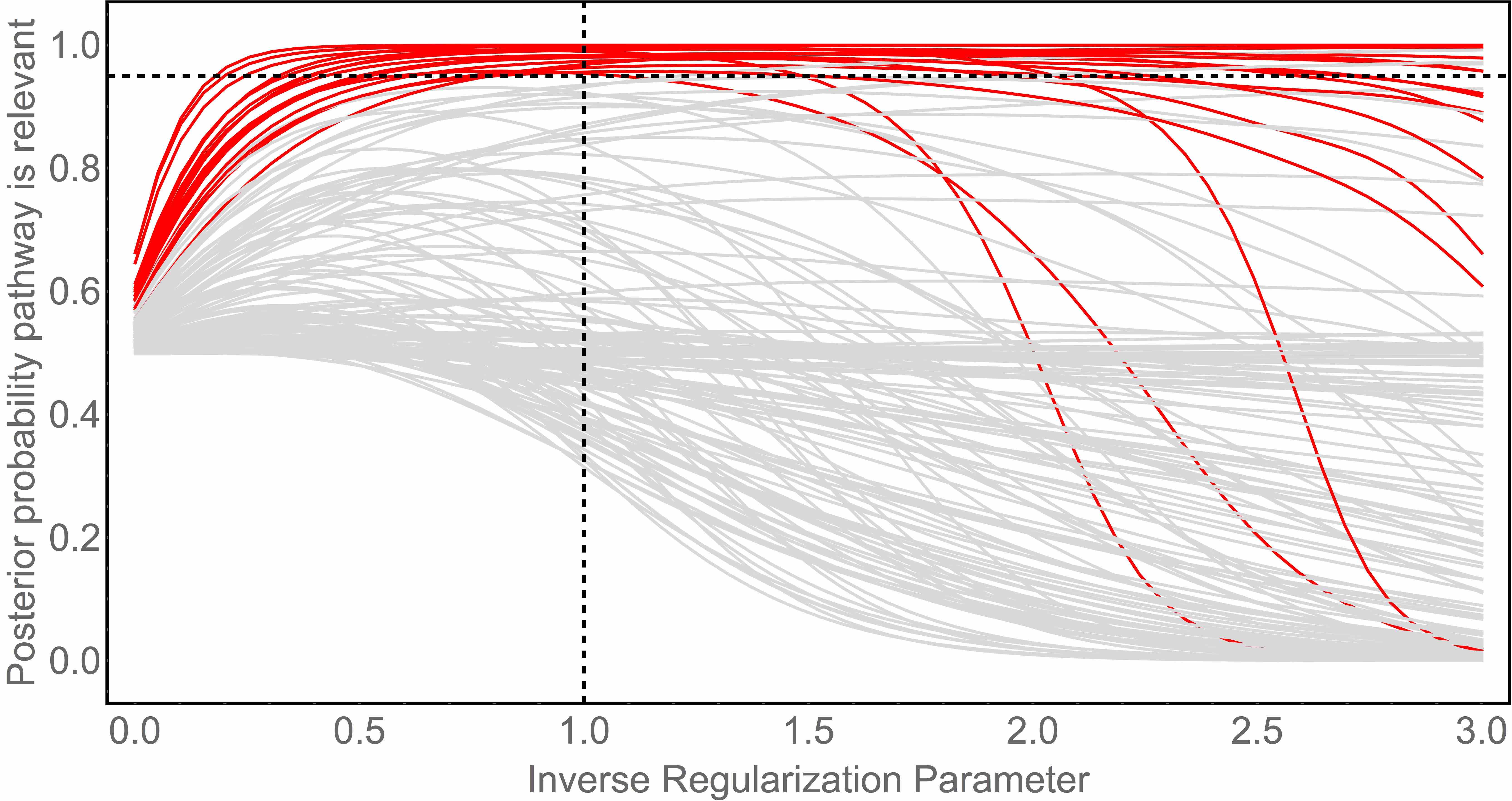}
\caption{ \textbf{Feature selection path of the Bayesian Ising Approximation.} Posterior probability that each figure (i.e., KEGG pathway) is relevant for computing the ecological distance between species as a function of the variance of the prior distribution (i.e., the inverse of the regularization parameter). The pathways with a posterior probability greater than $0.95$ when the inverse regularization parameter is one (i.e, $\lambda^*/\lambda$ = 1) are shown in red. \label{fig:figS8} }
\end{figure}

We have set up the problem so that identifying which pathways are relevant is equivalent to identifying those features for which $s_j = +1$. Therefore, we need to compute the posterior distribution for $\bvec{s}$, which can be determined from Bayes' theorem using $P_{\l}(\bvec{s} | \bvec{y}) \propto \int d\bvec{\beta} d\s^2 P(\bvec{y} | \bvec{\b}, \s^2) P_{\lambda}(\bvec{\b} ,\s^2 |\bvec{s}) P(\bvec{s})$. While the integral can be computed exactly, computing the marginal distributions (i.e. $P_{\lambda}(s_j | \bvec{y})$) is not computationally feasible. Therefore, we use an approach called the  Bayesian Ising Approximation (BIA). The BIA approximates the posterior distribution of the indicator variables using an Ising model described by:
\begin{align}
\log P_{\l}(\bold{s} | \bold{y}) 
&\simeq \frac{n^2}{4 \l}  \left(\sum_{i} h_i(\l) s_i + \HF \sum_{i,j;  i \neq j} J_{ij}(\l) s_i s_j \right)
\label{eq:ising}
\end{align}
where the external fields ($h_i$) and couplings ($J_{ij}$) are defined as:
\begin{widetext}
\begin{align}
h_i(\l) &= r^2(y,x_i) -\frac{1}{n} +\sum_{j} J_{ij}(\l) \label{eq:fields} \\
J_{ij}(\l) &= \l^{-1} r^2 (x_i,x_j) -  \frac{n}{\l} \left( r(x_i, x_j) r(y,x_i) r(y,x_j) - \HF  r^2(y,x_i) r^2(y,x_j) \right) \label{eq:couplings}
\end{align}
\end{widetext}
Here, $r(z_1,z_2)$ is the Pearson correlation coefficient between variables $z_1$ and $z_2$.  In writing this expression we have assumed that the hyperparameters $a_0$ and $b_0$ are small. The BIA approximation expansion converges as long as:
\be
\lambda \geq \lambda^* = n(1+ p r).
\ee
where $r =\sqrt{p^{-1}(p-1)^{-1} \sum_{i \neq j} r^2(X_i, X_j) }$ is the root mean square correlation between features. 

To perform feature selection,  we are interested in computing marginal probabilities $P_{\l}(s_j = 1 | \bold{y}) \simeq (1+ m_j(\l) )/2$, where we have defined the magnetizations $m_j(\l) = \< s_j \>$. While there are many techniques for calculating the magnetizations of an Ising model, we focus on the mean field approximation which leads to a self-consistent equation:
\be
m_i(\l) = \tanh \left[  \frac{n^2}{4 \l} \left( h_i(\l) + \HF \sum_{j\neq i}  J_{ij}(\l) m_{j}(\l) \right)  \right]  \label{eq:nmf}
\ee
This mean field approximation provides a computationally efficient tool that approximates Bayesian feature selection for linear regression, requiring only the calculation of the Pearson correlations and solution of Equation \ref{eq:nmf}.

As with other approaches to penalized regression, our expressions depend on a free parameter ($\l$) that determines the strength of the prior distribution. As it is usually difficult, in practice, to choose a specific value of $\l$ ahead of time it is often helpful to compute the feature selection path; i.e.\ to compute $m_j(\l)$ over a wide range of $\l$'s. Indeed, computing the variable selection path is a common practice when applying other feature selection techniques such as LASSO regression \cite{efron2004least}. To obtain the mean field variable selection path as a function of $\e = 1/\l$, we notice that $\lim_{\e \to 0} m_j(\e) = 0$ and so define the recursive formula:
\begin{widetext}
\be
m_i \left(\e +\d\e\right) \approx \tanh \left[  \frac{(\e + \d\e) n^2}{4} \left( h_i\left(\e + \d\e\right) + \HF \sum_{j \neq i} J_{ij}\left(\e + \d\e\right) m_j\left(\e\right) \right)  \right]  \nonumber
\ee
\end{widetext}
with a small step size $\d \e \ll 1/\l^{*} = n^{-1}( 1+ pr)^{-1}$. We have set $\d \e = 0.05 / \l^{*}$ in all of the examples presented below.

\begin{figure}
\centering
\includegraphics[width=\linewidth]{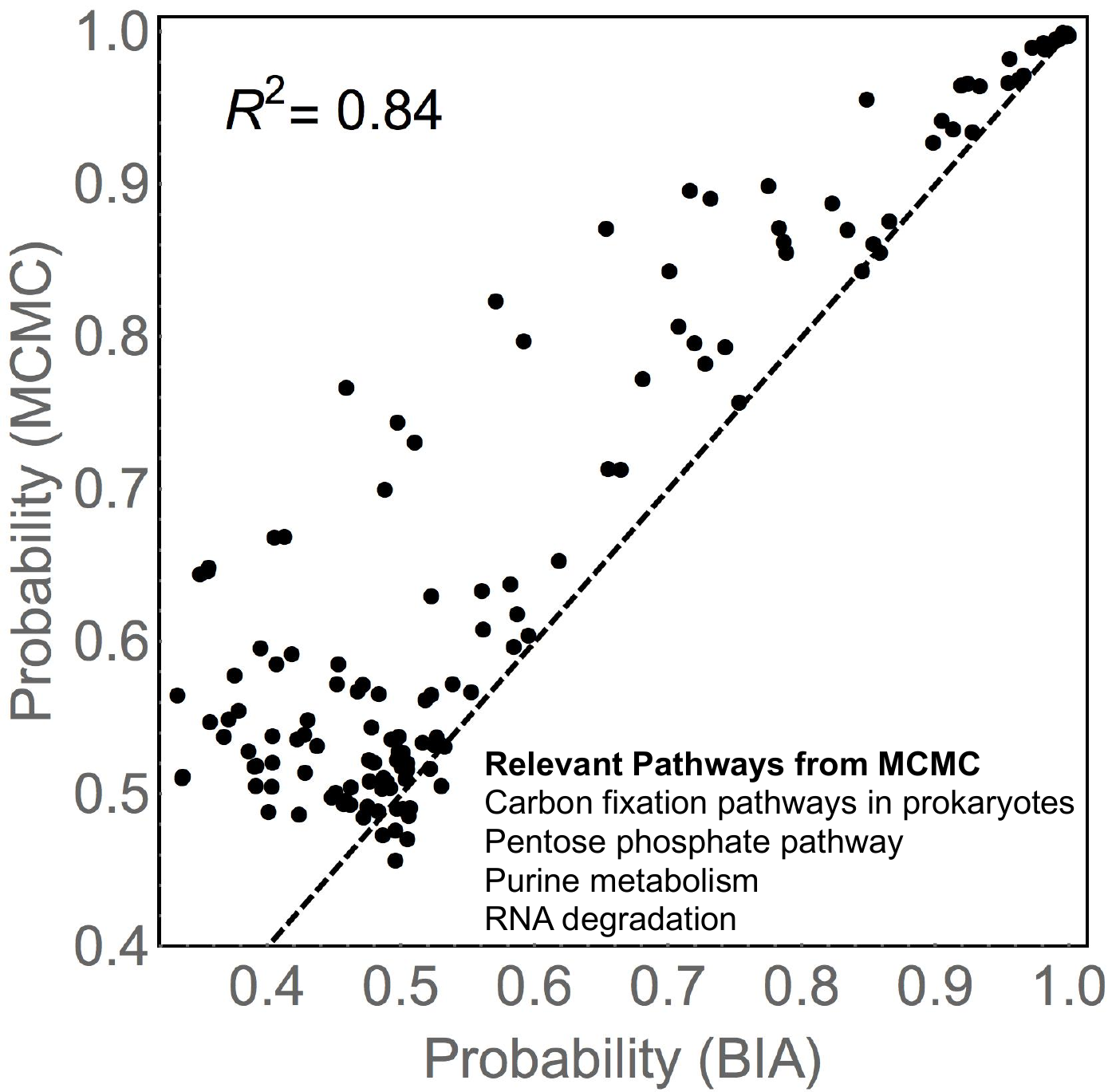}
\caption{ \textbf{Comparison of BIA to Monte Carlo simulations.} Posterior probabilities estimated using the BIA compared to those computed with Monte Carlo simulations for $\lambda = \lambda^*$. Four pathways (shown) reach a posterior probability of 0.95 for Monte Carlo, but not for BIA. All pathways that reached the 0.95 threshold for relevance with the BIA also reached the relevance threshold with Monte Carlo. \label{fig:figS9} }
\end{figure}

The feature selection path computed for the KEGG pathways using the BIA is shown in Figure \ref{fig:figS8}. This is a plot of $P_\lambda( \b_j \neq 0 | \matr{X})$ as a function of $\lambda^*/\lambda$. We focus on the point where $\lambda = \lambda^*$, which corresponds to the weakest prior distribution for which the BIA is applicable.  We defined pathways as significantly associated if they had a posterior probability greater than $0.95$. For comparison, posterior probabilities computed using Monte Carlo simulations of the exact posterior at $\lambda = \lambda^*$ are shown Figure \ref{fig:figS9}. The 17 significant pathways are: Aminoacyl-tRNA biosynthesis, Carbon fixation in photosynthetic organisms, Carbon metabolism, Citrate cycle (TCA cycle), Cysteine and methionine metabolism, Glutathione metabolism, Glycolysis/Gluconeogenesis, Homologous recombination, Lipoic acid metabolism, Lipopolysaccharide biosynthesis, Lysine biosynthesis, One carbon pool by folate, Porphyrin and chlorophyll metabolism, Pyrimidine metabolism, Pyruvate metabolism, Thiamine metabolism, and Vitamin B6 metabolism. A linear regression using just these relevant pathways has a correlation coefficient of $R^2 = 0.47$.

The BIA is, by definition, an approximation to the posterior probabilities. Although previous results suggest that the approximation is quite good \cite{fisher2015bayesian}, we performed Monte Carlo simulations of the exact posterior distribution with $\lambda = \lambda^*$ to validate that our conclusions were not highly sensitive to the approximation. A comparison of the results from BIA and Monte Carlo simulations is shown in Figure \ref{fig:figS9}. Four additional pathways (i.e., Carbon fixation pathways in prokaryotes, Pentose phosphate pathway, Purine metabolism, RNA degradation) achieve the $0.95$ significance threshold according to the Monte Carlo simulations, but all 17 pathways identified as relevant by the BIA were also identified as relevant by Monte Carlo. 

\section{Acknowledgements} 
We would like to thank Thomas Gurry for helpful conversations. This work was funded by a Philippe Meyer Fellowship to CKF. 

\bibliography{references}

\end{document}